\documentclass[12pt,preprint]{aastex}

\newcommand{\logg} {\log g}
\newcommand{\Te} {T_{\rm eff}}
\newcommand{\mv} {M_V}
\newcommand{\msun} {$M_\odot$}
\newcommand\gta{\lower 0.5ex\hbox{$\buildrel > \over \sim\ $}} 
\newcommand\lta{\lower 0.5ex\hbox{$\buildrel < \over \sim\ $}} 
\newcommand{\nh} {N({\rm H})/N({\rm He})}
\shortauthors{Limoges et al.}
\shorttitle{White Dwarfs from the Kiso Survey}
\begin{document}

\title{A Spectroscopic Analysis of White Dwarfs in the Kiso Survey}

\author{M.-M. Limoges and P. Bergeron}
\affil{D\'epartement de Physique, Universit\'e de Montr\'eal, C.P.~6128, 
Succ.~Centre-Ville, Montr\'eal, Qu\'ebec H3C 3J7, Canada.}
\email{limoges@astro.umontreal.ca, bergeron@astro.umontreal.ca}

\begin{abstract}
We present a spectroscopic analysis of white dwarfs found in the Kiso
survey. Spectroscopic observations at high signal-to-noise ratio have
been obtained for all DA and DB stars in the Kiso Schmidt ultraviolet
excess survey (KUV stars). These observations led to the
reclassification of several KUV objects, including the discovery of
three unresolved DA+DB double degenerate binaries. The atmospheric
parameters ($\Te$ and $\logg$) are obtained from detailed model
atmosphere fits to optical spectroscopic data. The mass distribution
of our sample is characterized by a mean value of 0.606 \msun\ and a
dispersion of 0.135 \msun\ for DA stars, and 0.758 \msun\ and a
dispersion of 0.192 \msun\ for DB stars. Absolute visual magnitudes
obtained from our spectroscopic fits allow us to derive an improved
luminosity function for the DA and DB stars identified in the Kiso
survey.  Our luminosity function is
found to be significantly different from earlier estimates based on
empirical photometric calibrations of $\mv$ for the same sample. The
results for the DA stars now appear entirely consistent with those
obtained for the PG survey using the same spectroscopic approach. The
space density for DA stars with $\mv\le12.75$ is $2.80\times10^{-4}$
pc$^{-3}$ in the Kiso survey, which is 9.6\% smaller than the
value found in the PG survey. The completeness of both surveys is
briefly discussed.

\end{abstract}


\section{Introduction}

White dwarf stars represent the final stage of stellar evolution for
main sequence stars whose masses lie between 0.07 and 8 \msun,
which correspond to about 97\% of the stars in the Galaxy. As the result of
the cessation of nuclear reactions, they simply cool off while
dissipating the content of their thermal reservoir. Because of these
characteristics, the white dwarf luminosity function --- the number of
white dwarfs as a function of their intrinsic luminosity --- is a
powerful tool that provides an estimate of the contribution of white dwarf stars to the
density of matter in the Galaxy. When the luminosity function is
derived from a complete sample of white dwarfs, it contains
information such as a direct measure of the stellar death rate in the
local galactic disk. The comparison with theoretical evolutionary
models \citep[see, e.g.,][]{fon01} then allows us to measure the age
of various components of the Galaxy. Likewise, the mass distribution
contains information about the amount of mass lost during the
evolution of an initial mass distribution \citep[][hereafter
LBH05]{LBH05}.

Luminosity functions have been determined for cool white dwarfs
discovered in high proper motion surveys, and through UV color-excess
surveys for hot white dwarfs. The cool end of the luminosity function
was studied by \citet{ldm88} and \citet{leggett98}, while the hot end
was analyzed by \citet{fleming86} using white dwarfs identified in the
Palomar-Green (PG) survey \citep{PG} and by Darling (1994; see also
\citealt{wegner94}) using the Kiso Schmidt ultraviolet excess survey
(KUV). LBH05 have recently improved upon the analysis of
\citet{fleming86} by applying the spectroscopic method
\citep{bergeron92} to measure the effective temperatures and surface
gravities of all the DA stars identified in the PG survey. The
improved atmospheric parameters allowed a better determination of the
absolute visual magnitude ($M_V$) of each star, and in turn improved
the accuracy of the luminosity function calculation. More recently, a
similar approach was applied by \cite{harris06} to a sample of 6000
white dwarfs (or white dwarf candidates) identified in the Sloan
Digital Sky Survey (SDSS) Data Release 3. The luminosity function
based on the SDSS extends to redder colors than the PG or Kiso
surveys, but does not extend to low enough temperatures to cover the
end of the white dwarf cooling sequence, although discoveries made
through proper motion diagrams may soon change this picture
\citep{kilic06}.

In order to measure the luminosity function of white dwarf stars, one
needs to carefully define a statistically complete sample. This is
a major endeavor, both for common proper motion or UV color excess surveys.
\citet{darling94} attempted to estimate the completeness of the PG and KUV 
surveys by counting the number of stars discovered in the overlapping
fields of both surveys (see also Section 4.4 below). LBH05 discuss at
length the completeness of the PG survey (their Section 4) by
comparing their improved luminosity function based on spectroscopic
$M_V$ values with the results of \citet{darling94} for the KUV survey
(see Figure 10 of LBH05). They evaluate the completeness of the PG
survey to 75\%, while \citet{darling94} found a lower value of 58\%.
However, this comparison between the PG and the Kiso surveys is
fundamentally flawed for at least two reasons. First, in the case of
the PG survey, only DA stars are considered while for the Kiso survey,
white dwarf stars of {\it all spectral types} are used in the
calculation of the luminosity function. Second, and most importantly,
the results of \citet{darling94} are based on $M_V$ values determined
from empirical photometric calibrations; this is similar to the
approach used by \citet{fleming86} for the PG survey. 

In this paper, we present a study aimed at improving the luminosity
function of white dwarf stars by applying the spectroscopic method to
the DA and DB stars identified in the Kiso survey. While our primary goal is
to improve the comparison of the luminosity functions derived from
the PG and Kiso surveys, we also provide an analysis of the global
properties of the KUV sample and demonstrate how samples of relatively bright
white dwarf stars still hide objects of significant
astrophysical interest. Our sample drawn from the Kiso survey is
presented in Section 2 and analyzed in Section 3 using the
spectroscopic technique. The absolute visual magnitudes obtained
from these accurate atmospheric parameter determinations are
then used in Section 4 to calculate an improved luminosity function
for these stars. A detailed comparison with the results of the
PG survey is also presented. Our conclusions follow in Section 5.

\section{Spectroscopic Content of the Kiso Survey}

The Kiso ultraviolet excess survey (KUV) is a photometric search for
UV-excess objects, performed with the 105-cm Schmidt telescope at the Kiso
Observatory. A total of 1186 objects were found in 44 fields in a belt
from the northern to the southern galactic pole at a galactic
longitude of $180\,^{\circ}$. The fields of the survey have a mean
limiting magnitude of $V=17.7$ and cover a total area of 1400 square
degrees \citep{noguchi80,kondo84}.

Our original goal was to secure optical spectroscopic observations for
all white dwarfs in the Kiso survey, regardless of their spectral
type.  Our starting point is the list of objects taken from Table 4.3
of \citet{darling94}. This list includes 234 white dwarfs identified
in the Kiso survey, which at that point in time was only 94\%
completed (a total of about 250 white dwarfs were expected to be found
after completion of the survey). Since cool white dwarfs are difficult
to analyze without photometric data, we decided to focus our
attention towards the hot end of the sample and exclude from our
spectroscopic survey all 15 cool white dwarfs of the DQ, DZ, and DC
spectral types. We thus retained a sample of 219 white dwarfs for
follow-up spectroscopic observations. Optical spectra for the majority
of the objects in this sample were obtained with the Steward
Observatory 2.3-m telescope equipped with the Boller \& Chivens
spectrograph and a Loral CCD detector. The 4.''5 slit together with
the 600 l mm$^{-1}$ grating in first order provided a spectral
coverage of $\lambda\lambda$3200--5300 at an intermediate resolution
of $\sim 6$~\AA\ FWHM. The spectra for the remaining southern objects
were obtained with the Carnegie Observatories' 2.5-m du Pont telescope
at Las Campanas (Chile), equipped with the Boller \& Chivens
spectrograph and a Tektronic (Tek 5) CCD detector.  The 1.''5 slit and
the 600 l mm$^{-1}$ grating provided a spectral coverage of
$\lambda\lambda$3500-6600 at a resolution of $\sim3$~\AA\ FWHM.
Exposure times were set to reach a signal-to-noise ratio of at
least $S/N\sim50$.

The better resolution and higher $S/N$ of our observations allowed us
to revise the spectral classification of many objects in this first
sample of 219 stars. At lower resolution, for instance, spectra of B
subdwarfs can be easily confused with a DA star (see LBH05 for several
examples). We found 30 objects in Darling's sample that turned out to
be subdwarfs or main sequence stars, 23 of which have been
reclassified as a result of our observations. These 30 objects are
displayed in Figure \ref{fg:f1} and reported in Table 1, together with
references to the original reclassification. Worth mentioning in this
figure is KUV 16032+1735, first identified as a magnetic DA white
dwarf by \citet{wegner90}. A closer examination of our spectrum
reveals no sign of a magnetic field, however, and an attempt to fit
this star with our DA model grid suggests an effective temperature
above 100,000~K. An independent analysis of this object, which will be
reported elsewhere, reveals that KUV 16032+1735 is actually a binary
system composed of a hot sdO star and a K-dwarf companion.  Additional
white dwarfs identified in the KUV survey were also observed as part
our spectroscopic follow-up. These are listed in Table 2. None of
these were part of the original data set used by \citet{darling94},
either because they had been discovered afterwards, or for reasons
unknown to us. For instance, KUV 14161+2255 (DB) had been reported by
\citet{wegner90} and should have been, in principle, part of Darling's
analysis.

Representative DA and DB/DO spectra from our survey are displayed in
Figures \ref{fg:f2} and \ref{fg:f3}, respectively, in order of
decreasing effective temperature, while magnetic DA white dwarfs are
displayed in Figure \ref{fg:f4}. The He~\textsc{ii} $\lambda4686$
absorption feature is clearly visible in the spectrum of the DAO star
KUV 03459+0037 shown in Figure \ref{fg:f2}. The spectrum for this star
was obtained from the SDSS Data Release Web
site.\footnote{http://www.sdss.org/dr7}Also shown in the same figure
are representative DA stars with an unresolved M-dwarf companion (KUV
01162$-$2311, KUV 03036$-$0043, KUV 04295+1739); we have a total of
six DA+dM systems in our survey. We also show in Figure \ref{fg:f3}
the spectrum of the DO star KUV 01018$-$1818 with a very strong
He~\textsc{ii} $\lambda4686$ line. Also observed is the presence of
H$\beta$ in several DB stars in our sample (DBA stars) as well
as Ca~\textsc{ii} H\&K lines (DBZ or DBAZ stars). Discussed further below are
three unresolved double-degenerate systems composed of DA+DB white
dwarfs that exhibit both hydrogen and helium lines. One of these
systems, KUV 02196+2816, has already been analyzed in detail by
\citet{lim09}. The results for the other systems, KUV 03399+0015 and
KUV 14197+2514, are presented in the next section.

Our final spectroscopic sample is thus the combination of the 219 hot
white dwarf candidates from \citet{darling94}, minus the 30
misclassified objects listed in Table 1, to which we add the 7
additional white dwarfs listed in Table 2, for a total of 196
spectroscopically confirmed white dwarfs, three of which are unresolved
double degenerates (thus a total of 199 white dwarfs). In summary, we
thus have 175 DA stars (including one DAO and 4 magnetics), 23 DB
stars, and one DO star.

\section{Atmospheric Parameter Determination}

Our model atmospheres and synthetic spectra for DA stars are described
at length in LBH05 and references therein. These are pure hydrogen,
plane-parallel model atmospheres. Non-local thermodynamic equilibrium
(NLTE) effects are explicitly taken into account above $\Te=20,000$~K
and energy transport by convection is included in cooler models
following the ML2/$\alpha=0.6$ prescription of the mixing-length
theory (see \citealt{B95}). The theoretical spectra are calculated
within the occupation formalism of \citet{HM88}, which provides a
detailed treatment of the level populations as well as a consistent
description of bound-bound and bound-free opacities. In order to
compare our results directly with those of LBH05, we refrain from
using here the new model spectra of \citet{tb09} that are based on
improved Stark profiles for the hydrogen lines; the full implications
of these improved models on the spectroscopic analysis of DA stars
will be presented elsewhere (Gianninas et al. 2010, in preparation).

Our model atmospheres and synthetic spectra for DB stars are similar
to those described in \citet{beauchamp96}, which include the improved
Stark profiles of neutral helium of \citet{beauchamp97}.  Our fitting
technique relies on the nonlinear least-squares method of
Levenberg-Marquardt \citep{press86}, which is based on a steepest
descent method. The model spectra (convolved with a Gaussian
instrumental profile) and the optical spectrum of each star are first
normalized to a continuum set to unity. The calculation of $\chi ^2$
is then carried out in terms of these normalized line profiles
only. Atmospheric parameters -- $\Te$, $\logg$, and $\nh$ for DBA
stars -- are considered free parameters in the fitting procedure. For
DA+dM binaries, H$\beta$ is simply removed from the $\chi^2$ fit when
the line is contaminated by the M-dwarf companion.

Atmospheric parameters for the 4 magnetic DA stars displayed in Figure
\ref{fg:f4} are taken from the literature: KUV 03292+0035
has a magnetic field of 12 MG \citep{jordan93} and an effective
temperature of $\Te\sim15,500$~K \citep{schmidt03}. The value of the
magnetic field for KUV 08165+3741 is 9 MG
\citep{angel74} with $\Te\sim 11,000$~K \citep{jordan01}. \citet{liebert85} report 29
MG for KUV 23162+1220 and $\Te\sim 11,000$~K based on {\it IUE}
data. Finally, KUV 23296+2642 has a magnetic field of 2.3 MG, with
$\Te=9400$~K and $\logg=8.02$ \citep{bergeron01}. For the first three
magnetic white dwarfs, we simply assume $\logg=8$.

The DA and DB white dwarfs concealed in DA+DB unresolved double degenerate
systems require a special treatment. The first such system discovered
in our survey, KUV 02196+2816, was spectroscopically identified as a
DBA star by \citet{darling96} but should have been classified as a DAB
star since the hydrogen lines are actually stronger than the helium
lines. \citet{lim09} showed that the optical spectrum of this object
cannot be reproduced by assuming model atmospheres with a homogeneous
hydrogen and helium composition, or even stratified atmospheres, and
that the hydrogen and helium lines observed in the spectrum of KUV
02196+2816 can be perfectly reproduced by assuming a double degenerate
binary composed of a DA star and a DB star. In this case, the predicted
spectrum is obtained by combining pure hydrogen and pure helium model
spectra, weighted by their respective radius. The detailed analysis
yields $\Te=27,170$~K and $\logg=8.09$ for the DA star, and
$\Te=36,340$~K and $\logg=8.09$ for the DB star. We identified
two additional systems in our spectroscopic survey, KUV 03399+0015 and
KUV 14197+2514.  KUV 03399+0015 was first identified as a DA? by
\citet{dw94}, and later as a DAB by \citet{eisen06}, while KUV
14197+2514 was classified as a DBA star by \citet{wegner90}. An
analysis similar to that of KUV 02196+2816 of these two objects
reveals that both spectra cannot be reproduced with single-star
models, and that the hydrogen and helium line profiles can only be
explained by assuming a composite DA+DB double degenerate model.
Figure \ref{fg:f5} summarizes our results for the three
systems. A more detailed analysis of these binaries will be presented
elsewhere (Limoges \& Bergeron 2010, in preparation).

The atmospheric parameters for the white dwarfs in our sample --- 192
single stars and 6 from the binary systems of Figure \ref{fg:f5} ---
are reported in Tables 3 and 4 for the DA and DB stars, respectively;
we exclude from our analysis the DO star KUV 01018$-$1818 since we
have no models to analyze this object properly. The values in
parentheses represent the uncertainties of each parameter, calculated
by combining the internal error obtained from the covariance matrix
and the external error estimated for DA stars at 1.2\% in $\Te$ and
0.038 dex in $\logg$ (see LBH05 for details); we assume the same
uncertainties for DB stars. The hydrogen abundances in Table 4 are
provided for DBA stars only, otherwise a pure helium composition is
assumed. The stellar mass ($M$) and the white dwarf cooling time
($\log \tau$, where $\tau$ is measured in years) are obtained from
detailed evolutionary cooling sequences appropriate for these
stars. For DA stars with $\Te> 30,000$~K, we use the carbon-core
cooling models of \citet{wood1995} with thick hydrogen layers of
$q({\rm H})\equiv M_{\rm H}/M_{\star}=10^{-4}$ and $q({\rm
He})=10^{-2}$, while for $\Te<30,000$~K, we use cooling models similar
to those described in
\citet{fon01} but with carbon-oxygen cores. For DB stars, we rely on
similar models but with thin hydrogen layers of $q({\rm H})=10^{-10}$
representative of helium-atmosphere white dwarfs. The absolute visual
magnitude ($\mv$) and luminosity ($L$) of each star are calculated
using the prescription of \citet{holberg06}\footnote{See
http://www.astro.umontreal.ca/\~{ }bergeron/CoolingModels.}, while the
apparent $V$ magnitudes are taken from the catalog of
\citet{ms06}. The $1/v_{\rm{max}}$ value corresponds to the maximum
volume in which it is possible to find a particular white dwarf given
a limiting apparent $V$ magnitude of the survey; this will be
discussed further in Section 3.5. Note that the $1/v_{\rm{max}}$ value
is provided only for white dwarfs in our statistically complete sample
defined below.

The global properties of the DA and DB stars in the KUV sample are
summarized in Figure \ref{fg:f6} in a mass versus effective
temperature diagram. This figure indicates that the majority of our DA
stars have effective temperatures between $\Te\sim10,000$~K and
25,000~K.  Only 13 stars in our sample have temperatures below
10,000~K; the coolest object is KUV 23235+2536 with $\Te=5930$~K.  An
examination of Figure \ref{fg:f6} reveals the well-known problem where
the spectroscopic masses for DA stars show a significant increase
below $\Te\sim13,000$~K (see also LBH05). Various explanations have
been proposed to account for this phenomenon
\citep{bergeron07,koester09}, although to this date none are
completely satisfactory. The most promising solution to this problem,
namely a mild and systematic helium contamination from convective
mixing that would mimic the high $\logg$ spectroscopic measurements,
has recently been ruled out by \citet{tremblay10} who reported
extremely low upper limits of the helium abundance in several cool DA
white dwarfs using high-resolution spectra from the Keck I 10-m
telescope. Figure \ref{fg:f6} also reveals that near $\sim15,000$~K, 4
DB stars have significantly higher masses than the rest of the DB
sample. This particular trend at low effective temperatures has been
observed in other analyses as well (see, e.g., Fig.~5 of
\citealt{kepler07} for the DB stars identified in the SDSS) and this
problem is usually attributed to uncertainties in the treatment of van
der Waals broadening of neutral helium lines \citep{beauchamp96}. In
the results shown here, however, DB stars with normal masses ($\sim
0.6$ \msun) are also found in the same range of temperatures. To
illustrate this more clearly, we show in Figure \ref{fg:f7} our best
fits to two DB stars near 15,000~K, but with significantly different
$\logg$ values. In this particular range of temperatures, the
He~\textsc{i} $\lambda3820$ and $\lambda4388$ lines are the most
gravity sensitive and the strengths of these lines do indeed appear
very different in both stars. We emphasize that this result for DB
stars is completely independent of any modeling of the line profiles
and we must therefore conclude that the spread in mass observed in
Figure \ref{fg:f6} may thus be real after all. Perhaps the massive DB
stars observed here represent the descendants of the new class of
white dwarf stars with carbon-rich atmospheres discovered by
\citet{dufour07,dufour08}. Indeed, the coolest `hot DQ' stars have
temperatures near $\sim 18,000$~K, below which they are presumably
transformed into something else. Small amounts of helium thoroughly
mixed in their atmospheres could easily reappear at the surface of
these stars below 18,000~K or so, turning them into DB stars in a
process similar to the transformation of hot DO (or even DAO) stars
into DA stars near $\sim45,000$~K.

The mass distribution is shown in Figure \ref{fg:f8} for the 136 DA
stars above 13,000~K (i.e., above the temperature where the masses of
DA stars are reliable) as well as for the 23 DB white dwarfs. The mass
distribution of DA stars has a mean value of 0.606 \msun\ with a
dispersion of $0.135$ \msun. These values are very similar to those
obtained by LBH05 for the DA stars in the PG sample (0.603 and 0.134
\msun, respectively) in the same temperature range. The mean mass for 
the DB stars in the KUV sample is 0.758 \msun, a value significantly
larger than that for the DA stars; the dispersion is also much larger.
The mass distribution for the DB stars appears rather flat compared to
those presented in \citet[][54 DB stars]{beauchamp96} or in
\citet[][71 DB stars]{voss07}, although the number of DB stars in our
sample is admittedly smaller than in these two previous studies. We
note finally that we found no low mass ($M<0.5$~\msun) DB white dwarfs
in our sample, in agreement with the results of \citet{beauchamp96}.

\section{Luminosity Function}

\subsection{General Considerations}

We can now proceed to calculate the luminosity function of DA and DB 
stars in the Kiso survey. Since we will compare our results with those of
\citet{darling94} and with the PG survey (LBH05), it is relevant to 
provide here some details of how these calculations are performed. The
luminosity function is obtained following the $1/v_{\rm{max}}$ method,
originally developed by \citet{schmidt68} in the context of quasars.
This method requires that we determine for each star the maximum
volume, $v_{\rm{max}}$, in which the object would have been detected
given the limiting magnitude of the survey. The distance to each star
can be directly obtained from the apparent and absolute $V$
magnitudes. Since the stars are not uniformly distributed in the
direction perpendicular to the galactic disk, but instead follow an
exponential disk, we define a weighted volume
$dv^\prime=\exp(-z/z_0)\,dv$, where $z=r\sin\theta$ is the distance of
the object from the galactic plane, $z_0$ is the galactic disk scale
height, and $\theta$ is the absolute value of the galactic latitude of
the object. Here we assume $z_0=250$~pc as in LBH05.

The Kiso survey is a magnitude-limited survey, which implies that it
must be complete down to a given limiting magnitude
$V_{\rm{lim}}$. For a given star with a magnitude $V$, $V_{\rm{lim}}$
defines a maximum distance $d_{\rm max}$ at which an object could
have been observed and still have been found in the survey. This in turn
defines the maximum volume, $v_{\rm{max}}$. As mentioned in
\citet{green80}, however, each photographic plate has a different limiting
magnitude. It was then proposed to define a maximum volume for each
field. The total maximum volume for a given star is thus the sum of
all the small individual volumes. Also, in the
case where two fields or more overlap, the area of overlap is given to
the field with the fainter limiting magnitude. We finally obtain
for each star

\begin{equation}
v_{\rm{max}}={\sum_{j=1}^{n_f}}\,\frac{\omega_j}{4\pi}\int_{0}^{d_{\rm max}}e^{-z/z_0}~4{\pi}r^2dr
\end{equation}

\noindent
where ${n_f}$ is the total number of fields. Since $\omega_j$ represents
the area covered by each field, $\sum_{j=1}^{n_f}\omega_j$ must be equal
to the total area covered by the survey (in steradians). The limiting
magnitude can be calculated following the $v/v_{\rm{max}}$ test of
uniformity of \citet{schmidt68}. The method states that a sample is
complete when the average value of $v/v_{\rm{max}}$ is equal to 0.5.
Once the value of $V_{\rm{lim}}$ that satisfies this constraint has
been found, all white dwarfs with $V$ fainter than $V_{\rm{lim}}$ must
be removed from the sample. The remaining stars define the complete
sample from which the luminosity function can be derived.

Since $v_{\rm{max}}$ represents the volume within which it is possible to
find a particular white dwarf, the contribution of each star to
the local space density is then simply given by $1/v_{\rm{max}}$
\citep{schmidt68}. The differential luminosity function as a function of $\mv$ can
finally be obtained by summing all the individual contributions to
the local space density:

\begin{equation}
\phi(\mv)={\sum_{i=1}^{n_b}}\ 1/v_{{\rm max}_i}
\end{equation}

\noindent
where $n_b$ is the number of stars in each magnitude bin of the
complete sample.  The uncertainties are evaluated as if the stars were
distributed randomly in space. The statistical uncertainty in each bin
is then given by $\sigma_\phi=[\sum_{i=1}^{n_b}{(1/v_{{\rm
max}_i})^2}]^{1/2}$ \citep{boyle89}.

\subsection{The Luminosity Function from Darling (1994)}

In the next section, we will compare our luminosity function of DA 
and DB stars based on spectroscopic estimates of $\mv$ with the results of
\citet{darling94} based on empirical photometric calibrations of
$\mv$.  Prior to this, we must first demonstrate that we are able to
reproduce the luminosity function displayed in his Figure 4.2 (also
reproduced in Figure 10 of LBH05) using the $V$ and $\mv$ values
provided in his Table 4.3. Note in this case that the luminosity
function is calculated using all 234 white dwarf stars identified in
the Kiso survey, all spectral types included. His method is identical
to that described above with the only exception that there is no
explicit sum over all fields in equation (1). Instead, the area of
each field is assumed to be constant, in which case a single factor
$\omega/4\pi$ appears in the equation, where $\omega$ for the Kiso
survey is 1400 square degrees (or $\sim 0.43$ steradians). We know
that in \citet{wegner94}, the luminosity function was scaled by a
certain factor to account for the fact that the Kiso survey was not
complete at that time. In \citet{darling94}, this factor had a value
of $1186/1115$ since the Kiso survey was only 94\% complete. This
correction factor is included in all our calculations below.

The comparison of the results between Darling's calculations and ours
using the same stars, as well as $V$ and $\mv$ values, is displayed in Figure
\ref{fg:f9}. Both functions have the same exact shape but ours
is scaled downwards by a factor of $\sim 2.5$, which suggests a
constant correction factor between both calculations. We also note
that our calculation of the limiting magnitude of the survey using
this sample agrees perfectly with the value obtained by Darling,
$V_{\rm lim}=17.35$. As it is not stated explicitly in
\citet{darling94} that his luminosity function contains a
correction factor (to account for other types of incompleteness, for
instance), we ignore the source of the difference between our
calculations and Darling's and use our method of calculation in the
remainder of this paper\footnote{We also contacted G.W.~Darling
directly and attempted to figure out the discrepancy but the original
calculations are lost in the sands of time.}. We must finally
mention that the code we are using here is the same as that used by
LBH05 so the comparisons discussed below will be entirely consistent.

\subsection{An Improved Luminosity Function for KUV White Dwarfs}

The main goal of our spectroscopic survey is to improve upon the
luminosity function of \citet{darling94} by using spectroscopic values
of $\mv$ rather than empirical photometric calibrations. The absolute
magnitudes are central to the calculation of the luminosity function
and it is crucial to measure $\mv$ values as accurately as possible.
In \citet{darling94}, $\mv$ values were assigned for 125 stars (out of
234) with measured $(B-V)$ using empirical $(B-V)$ vs.~$\mv$ relations
derived by \citet{sion77} and \citet{dahn82} for hot and cool white
dwarfs, respectively. Additional $\mv$ values for 9 objects were
directly taken from \citet{ms87}. Finally, a linear fit based on
measured photometric $(B-V)$ and photographic $(m_U-m_G)$ color
indices was then used to assign approximate $(B-V)$ values, and
thus $\mv$ values, to the remaining 100 stars without $(B-V)$ or $\mv$
measurements. In reality, this relation is far from being linear and a
substantial dispersion is present in the data (see Figure 4.1 of
\citealt{darling94}), potentially introducing a significant uncertainty 
in the $\mv$ estimates.

Our spectroscopic $\mv$ values taken from Tables 3 and 4 are compared
in Figure \ref{fg:f10} with those derived by \citet{darling94} for the
192 white dwarf stars in common. In doing so, we explicitly excluded from
Darling's sample all objects that have been spectroscopically
misclassified in the Kiso survey (Table 1 and Figure \ref{fg:f1}).
The double degenerate binary components are compared with Darling's
results for a single DAB/DBA star. The differences observed are quite
significant, with the empirical estimates being generally fainter than
the spectroscopic values. In some cases, the differences can reach 2
to 3 magnitudes. It is therefore expected that the corresponding
luminosity functions based on these absolute magnitudes will be
affected as well.

With the spectroscopic $\mv$ values for our sample of 175 DA and 23 DB
stars provided in Tables 3 and 4 respectively, we can now proceed to
the evaluation of the luminosity function. The first step is to find
the limiting magnitude of the survey following the $v/v_{\rm{max}}$
method described above. We find $V_{\rm lim}=17.41$, which defines a
complete sample of 168 white dwarf stars (149 DA and 19 DB
stars). This complete sample is flagged in Tables 3 and 4 by a value
of $1/v_{\rm{max}}$ in the appropriate column (otherwise the field is
empty). The luminosity function is then simply computed by summing
these values in appropriate bins (see equation 2), in this case
half-magnitude bins.

We first compare our improved luminosity function with that of
\citet{darling94} by using the same subset of 192 stars shown in
Figure \ref{fg:f10}, (i.e.~misclassified objects from Darling's sample
and the additional white dwarfs from Table 2 are not taken into
account). Even though the stars used in the comparison are the same,
the absolute magnitudes in each sample are evaluated differently, and
so are the limiting magnitudes. We find using Darling's $\mv$ values a
limiting magnitude of $V_{\rm lim}=17.35$, which defines a complete
sample of 161 objects, while our spectroscopic $\mv$ values yield
$V_{\rm lim}=17.41$ for a complete sample of 164 objects. The
corresponding luminosity functions are compared in Figure
\ref{fg:f11}. Despite the large differences in $\mv$ values observed
in Figure \ref{fg:f10}, the differences in the luminosity functions do
not appear as significant, with the noticeable exception of the 9.5,
12.5, and 13.0 magnitude bins. A similar conclusion was reached by
LBH05 when comparing the luminosity function of PG stars with the
earlier estimates of \citet{fleming86}. We must note that the main
effect here is to shift the number of stars from one bin to
another. For instance, the largest difference observed is near the
peak of the luminosity function where our determination based on
spectroscopic $\mv$ values is lower than that of Darling.  Indeed, a
lot of the objects in these magnitude bins in Darling's analysis have
been shifted to brighter bins as a result of his overestimates of
$\mv$ values (see Figure~\ref{fg:f10}). We finally note that the last
two bins at $\mv=14.0$ and 14.5 are populated by only one star each
(KUV 23235+2536 and KUV 08275+3252, respectively). The Kiso survey is
obvioulsy very imcomplete in this region and we decided to remove
these two white dwarfs from our sample in the remainder of our
analysis\footnote{KUV 23235+2536 has an estimated temperature of only
$\Te=5930$~K, and it is difficult to understand how such a cool
object ended up in a UV color-excess survey!}.

In Figure 10 of LBH05, the authors compare their luminosity function
for DA stars in the PG survey with the results of Darling (1994),
which, as mentioned above, also contains the contributions from other
spectral types. In order to improve this comparison, we derived an
independent KUV luminosity function for DA stars only, by simply
removing the DB white dwarfs from our sample (DB stars account for
17.3\% of our complete sample). It is now possible to make a detailed
comparison with the luminosity function obtained by LBH05, which is
based on the exact same analysis, including model atmospheres,
observing setup, fitting technique, and luminosity function
calculations. As mentioned explicitly in LBH05, their luminosity
function did not undergo any correction, as is the case with ours. It
is then possible to compare the results from both functions directly,
as shown in Figure~\ref{fg:f12}. The first result is that both
luminosity functions agree much better than estimated by LBH05, who
concluded, based on the results shown in their Figure 10, that ``the
KUV luminosity function has significantly more stars than PG in the
bins $\mv$ 10.5 to 13.0, with the PG appearing to become incomplete by
a factor of 4 in the 12.0 and the two fainter magnitude bins''. The
comparison shown here indicates on the contrary that both functions
agree perfectly within the error bars in these particular bins. 
However, there appears to be a deficiency of luminous
white dwarfs ($\mv\leq10$) in the Kiso survey. It is
perhaps not surprising that the luminosity functions from the PG and
Kiso surveys are so similar, since the detection method is the same
(UV excess). So whatever incompleteness factors affect the PG survey
seem to affect the Kiso survey as well.

Another way to compare both luminosity functions is to calculate the
local space densities of white dwarfs, which can be simply obtained by
integrating the luminosity function over a given range of $M_V$. For
$\mv<12.75$, we obtain $2.80\times10^{-4}$ ${\rm pc}^{-3}$ while LBH05
report a value\footnote{LBH05 actually give a value of
$5.0\times10^{-4}$ ${\rm pc}^{-3}$ but this number is inaccurate for
reasons unknown to one of the co-authors of both studies (P.B.) who
obtained the correct number provided here.} of $3.07\times10^{-4}$ ${\rm
pc}^{-3}$ in the same magnitude interval, or a value only 9.6\% larger
than our estimate.  The Kiso survey thus appears to be {\it less
complete}, if anything, than the PG survey in that range of absolute
visual magnitude. The Kiso survey is certainly deeper than PG, and it
should thus be more complete for fainter magnitude bins since the PG
survey is known to be fairly incomplete at the faint end of the
luminosity function (LBH05). We note, however, that the differences do
not appear as large as previously estimated by LBH05, as discussed
above.

Finally, our final luminosity function for the Kiso white dwarfs,
including all 149 DA and 19 DB stars in the complete sample, is
displayed in Figure \ref{fg:f13}. We obtain from the sum of all
magnitude bins a local space density of white dwarfs in the Kiso
survey of $5.49\times10^{-4}$ ${\rm pc}^{-3}$. Of course, the Kiso
sample does not extend to the faint magnitudes where the peak of the
luminosity function occurs ($\mv\sim15.5$), and the space density
determined here accounts for only $\sim 11$\% or less of the total
space density of white dwarf stars, which is estimated at
$5\times10^{-3}$ ${\rm pc}^{-3}$ or so.  For completeness, we also
compare in Figure \ref{fg:f14} our luminosity function with the
results obtained by \citet{harris06} and \citet{degen08} for the white
dwarfs discovered in the SDSS (Data Release 3). Our luminosity
function is plotted here as a function of $M_{\rm bol}$ to be
consistent with the SDSS determinations. Our results agree fairly
well, within the uncertainties, with the results of DeGennaro et
al.~for magnitude bins between $M_{\rm bol}=7.5$ and 11.5 (with a few
notable exceptions), but appear seriously underestimated for brighter and
fainter magnitude bins.

\subsection{Completeness of the PG and Kiso Surveys}

By counting the number of KUV white dwarfs (all spectral types
included) found by the PG survey in the overlapping fields (600 square
degrees of overlap) and within the magnitude limit of the PG survey,
\citet{darling94} estimated the completeness of the PG survey to
57.5\%, or 60.5\% if white dwarfs not in the statistically complete PG
sample are also included. However, as discussed in Section 2, many KUV
white dwarfs turn out to be lower gravity objects. A similar estimate
of the completeness of the PG survey based on our spectral
reclassification of DA and DB stars yields instead a value of 67.6\%.

\citet{darling94} also used the PG survey to estimate the completeness 
of the Kiso survey by turning the problem around, and by counting the
number of PG stars in the overlapping fields, all spectral types
included, that were actually missed in the Kiso survey. While the Kiso
survey is much deeper ($V_{\rm lim}\sim 17.35$) than the PG survey
($V_{\rm lim}\sim 16.16$), only 74.2\% of the white dwarfs in the
overlapping fields of the PG survey were recovered in the Kiso
survey! Since none of the new KUV stars (Table 2) are part of the PG
sample, this estimate is still valid. Once again, using only our
spectroscopically confirmed sample of DA and DB stars, we find instead a
value of 84.4\% for the completeness of the Kiso survey. Given the
results shown in Figure \ref{fg:f12}, it is likely that both the PG
and Kiso surveys suffer from a comparable level of incompleteness.  We
refer the reader to the additional discussion of LBH05 (Section 4)
regarding the completeness of the PG survey. Obviously, deeper and
more complete surveys are still badly needed.

The Sloan Digital Sky Survey (SDSS) is certainly among the most
important developments in the last few years in terms of observational
data of white dwarf stars since the PG survey. The SDSS is looking at
10,000 deg$^2$ of high-latitude sky in five bandpasses ($ugriz$) and
is producing images in these five bandpasses from which galaxies,
quasars, and stars are selected for follow-up spectroscopy.  The
selection effects in this survey are important, as discussed in
\citet{kleinman04} and \citet{eisen06}, and therefore, it cannot be considered 
as a complete survey in any sense. For instance, \citet{degen08} found
a completeness of 51\% for their uncorrected sample of white dwarfs
from the SDSS, which is mainly composed of DA stars.  Also, the $S/N$
of the SDSS spectra is known to be proportional to the brightness of the
object, since the integration time is fixed. This can lead to large
uncertainties in the determination of the atmospheric parameters of
fainter objects \citep{gianninas05}. Finally, as mentioned in
\citet{eisen06}, ``completeness is not our goal'', since they know that
white dwarfs with $\Te\ \lta8000$~K are lost because of the color-cut, and
that magnetic white dwarfs can pass through the detection system
without being noticed. \citet{degen08} are in possession of a much
larger sample than ours, which allows them to divide their luminosity
function into several mass components.  However, when conducting a
statistical analysis of a sample, the completeness of this sample
should be a crucial parameter if the results are to be interpreted
physically.

\section{Conclusion}

We presented an analysis of the DA and DB white dwarfs in the KUV
survey, and determined the atmospheric parameters for each star from
detailed model atmosphere fits to optical spectroscopic data. The
$\mv$ values derived from the atmospheric parameters were compared
with those of \citet{darling94}, which were obtained from photometric
empirical calibrations. Our study allowed us to measure directly the
impact of the use of state-of-the-art model atmospheres on the
determination of absolute magnitudes for white
dwarfs. The differences were found to be significant, but had a
somewhat smaller impact on the calculation of the luminosity function.

We then proceeded to derive the luminosity function of DA and DB stars
found in the Kiso survey. We find as a result of our improved $M_V$
values a smaller number of stars in the fainter magnitude bins than
estimated by \citet{darling94}.  The comparison of our luminosity
function with that of LBH05, for DA stars only, reveals that both
functions are similar. We obtained a local space density of white
dwarfs of $5.49\times10^{-4}$ $\rm{pc}^{-3}$, while this number drops
to $2.80\times10^{-4}$ $\rm{pc}^{-3}$ for $\mv\le12.75$. These results
are now entirely consistent with those published in LBH05 for the PG
survey and the completeness of 
both surveys appears comparable.

Our spectroscopic survey of white dwarfs in the Kiso survey has also
led to an important spectral reclassification of the sample published
in \citet{darling94}.  In particular, we have identified three
unresolved double degenerate binaries. A two-component fit confirmed
that KUV 02196+2816, KUV 03399+0015, and KUV 14197+2514 are unresolved
double degenerate binaries composed of a DA and a DB star. These
systems were easily identified in our analysis because both components
had different spectral types. However, double degenerates composed of
two DA stars would go totally unnoticed, as demonstrated by
\citet{liebert91}. One may wonder how many such binaries might be
hiding in spectroscopic surveys such as KUV or PG. This in turn could
affect our determination of the true space density of white dwarf
stars.

This era of large scale surveys where the samples can contain up to
thousands of stars will certainly help us in the characterization of
our Galaxy. Accurate statistical analyses will then provide even more
precise determinations of, for example, the white dwarf space
densities in the different populations of the Galaxy, the stellar
contribution to the mass of the Galaxy, the age of the local galactic
disk, the stellar formation and death rates, etc. The issue of
completeness is thus of great importance when statistical analyses of
these samples are considered.

We would like to thank the director and staff of Steward Observatory
and Carnegie Observatories for the use of their facilities. We would also
like to thank A.~Gianninas for the acquisition of the spectra in the
southern hemisphere and for a careful reading of this manuscript. This
work was supported in part by the NSERC Canada and by the Fund FQRNT
(Qu\'ebec). P.B. is a Cottrell Scholar of Research Corporation for
Science Advancement.

\clearpage

\clearpage
\begin{deluxetable}{llccr}
\tabletypesize{\scriptsize}
\tablecolumns{12}
\tablewidth{0pt}
\tablecaption{Misclassified Objects}
\tablehead{
\colhead{KUV} &
\colhead{WD} &
\colhead{ST (Darling 1994)} &
\colhead{ST (this work)} &
\colhead{Notes}}
\startdata
00486$-$2016   &0048$-$202     &DA   &sdB  &1       \\
01098$-$2629   &0109$-$264     &DA   &sdB  &1       \\
01134$-$2423   &0113$-$243     &DA   &sdB  &2       \\
01542$-$0710   &0154$-$071     &DA?  &sdB  &1       \\
02222+3124     &0222+314       &DA   &sdB  &2       \\
02409+3407     &0240+341       &DA   &MS   &2       \\
04473+1737     &0447+176       &DB   &sdOB &2       \\
05097+1649     &0509+168       &DA   &MS   &3       \\
05101+1619     &0510+163       &DA   &MS   &2       \\
05260+2711     &0526+271       &DA   &sdB  &2       \\
05296+2610     &0529+261       &DA   &sdB  &2       \\
06274+2958     &0627+299       &DA   &sdB  &2       \\
06289+3126     &0628+314       &DA   &sdB  &2       \\
07528+4113     &0752+412       &DA?  &sdB  &2       \\
09272+3854     &0927+388       &DA   &sdB  &2       \\
09306+3740     &0930+376       &DA   &sdB  &2       \\
09327+3937     &0932+396       &DA   &sdB  &2       \\
09339+3821     &0933+383       &DA   &sdB  &4      \\
09372+3933     &0937+395       &DA   &sdB  &2       \\
09436+3709     &0943+371       &DA   &sdB  &2       \\
09467+3809     &0946+381       &DA   &sdB  &2       \\
12562+2839     &1256+286       &DA?  &sdB  &2       \\
13024+2824     &1302+284       &DA   &sdB  &4, 5  \\
13023+3145     &1302+317       &DB   &sdOB &2       \\
13046+3118     &1304+313       &DA   &sdB  &2       \\
16032+1735     &1603+175       &DA   &sdO+dK  &2       \\
16118+3906     &1611+390       &DA   &sdB  &2       \\
18169+6643     &1816+667       &DB   &sdOB &2       \\
22585+1533     &2258+155       &DB   &sdO  &6      \\
23099+2548     &2309+258       &DA   &sdB  &2       \\
\enddata
\tablecomments{
Note. $-$ (1) \citealt{lis05}; 
(2) this work; 
(3) \citealt{kaw04};
(4) Not in  McCook and Sion; 
(5) \citealt{holb03}; 
(6) \citealt{stroeer07}.}
\end{deluxetable}

\clearpage
\begin{deluxetable}{llcr}
\tabletypesize{\scriptsize}
\tablecolumns{12}
\tablewidth{0pt}
\tablecaption{Additional KUV White Dwarfs}
\tablehead{
\colhead{KUV} &
\colhead{WD} &
\colhead{ST} &
\colhead{Reference}}
\startdata
01018$-$1818&0101$-$182&DO&\citet{lamont00}\\
03439$-$0048&0343$-$007&DA&\citet{wagner} \\
03459+0037&0345+006&DAO&\citet{eisen06}\\
08422+3813&0842+382&DA&\citet{kaw05}\\
14161+2255&1416+229&DB&\citet{wegner90}\\
18004+6836&1800+685&DA&\citet{kidder91}\\
18453+6819&1845+683&DA&\citet{kidder91}\\
\enddata
\end{deluxetable}

\clearpage
\begin{deluxetable}{llrrlcrclccclc}
\tabletypesize{\scriptsize}
\tablecolumns{12}
\tablewidth{0pt}
\rotate
\tablecaption{Atmospheric Parameters of DA Stars from the KUV Sample}
\tablehead{
\colhead{KUV} &
\colhead{WD} &
\colhead{$\Te$} &
\colhead{(K)} &
\colhead{log $g$} &
\colhead{$M/M_{\odot}$} &
\colhead{$M_V$} &
\colhead{log $L/L_{\odot}$} &
\colhead{$V$} &
\colhead{$D ({\rm pc})$} &
\colhead{$1/v_{\rm max}$} &
\colhead{log $\tau$} &
\colhead{Notes}}
\startdata
 00300$-$1810&   0030$-$181  & 13,640   &(362) &7.81 (0.06)    &0.51 (0.00)    &11.18          &$-$2.17        &16.8        &  133&1.78($-$6)     &8.30           &                    \\
 00328$-$1735&   0032$-$175  &  9830    &(142) &8.18 (0.05)    &0.71 (0.00)    &12.53          &$-$2.97        &14.9        &   29&9.92($-$6)     &8.91           &                    \\
 00329$-$1747&   0032$-$177  & 16,780   &(281) &7.83 (0.05)    &0.52 (0.00)    &10.83          &$-$1.82        &15.7        &   94&1.16($-$6)     &8.00           &                    \\
00334$-$1738 &  0033$-$176   & 23,030   &(477) &7.98 (0.06)    &0.62 (0.00)    &10.49          &$-$1.35        &17.6        &  264&            ---&7.51           &                    \\
00337$-$1749 &  0033$-$178   & 24,310   &(572) &7.26 (0.08)    &0.34 (0.00)    & 9.25          &$-$0.79        &17.7        &  489&            ---&7.24           &                    \\
\\
 00442$-$2156&   0044$-$219  & 11,890   &(207) &7.84 (0.07)    &0.51 (0.00)    &11.47          &$-$2.43        &17.26       &  143&2.55($-$6)     &8.49           &                    \\
 00582$-$1834&   0058$-$185  & 17,760   &(335) &8.02 (0.06)    &0.63 (0.00)    &11.02          &$-$1.83        &17.26       &  177&1.46($-$6)     &8.07           &                    \\
 01024$-$1836&   0102$-$185  & 72,370   &(1793)&7.16 (0.08)    &0.48 (0.00)    & 7.13          &$+$1.35        &16.9        &  900&2.61($-$8)     &5.07           &                    \\
 01071$-$1917&   0107$-$192  & 14,750   &(254) &7.83 (0.05)    &0.52 (0.00)    &11.06          &$-$2.04        &16.3        &  111&1.53($-$6)     &8.20           &                    \\
01138$-$2431 &  0113$-$245   & 56,370   &(1975)&7.58 (0.13)    &0.54 (0.00)    & 8.29          &$+$0.55        &17.67       &  751&            ---&6.15           &                    \\
\\
 01157$-$2546&   0115$-$257  & 14,990   &(264) &7.94 (0.05)    &0.58 (0.00)    &11.19          &$-$2.08        &16.90       &  138&1.80($-$6)     &8.25           &                    \\
 01162$-$2311&   0116$-$231  & 31,430   &(489) &7.65 (0.06)    &0.49 (0.00)    & 9.30          &$-$0.57        &15.63       &  184&1.98($-$7)     &6.90           &1                   \\
 01246$-$2546&   0124$-$257  & 24,110   &(392) &7.73 (0.05)    &0.50 (0.00)    &10.03          &$-$1.11        &15.66       &  133&4.48($-$7)     &7.28           &                    \\
 01308$-$2721&   0130$-$273  & 22,490   &(376) &7.92 (0.05)    &0.58 (0.00)    &10.44          &$-$1.35        &16.93       &  198&7.24($-$7)     &7.50           &                    \\
 01491$-$1127&   0149$-$114  &  9120    &(137) &8.03 (0.07)    &0.62 (0.00)    &12.59          &$-$3.01        &16.6        &   63&1.07($-$5)     &8.90           &                    \\
\\
 01498$-$1227&   0149$-$124  & 12,180   &(199) &8.28 (0.05)    &0.78 (0.00)    &12.05          &$-$2.65        &17.2        &  106&5.33($-$6)     &8.73           &                    \\
01518$-$0928 &  0151$-$094   & 26,000   &(460) &7.75 (0.06)    &0.51 (0.00)    & 9.89          &$-$0.99        &17.7        &  364&            ---&7.16           &                    \\
 01552$-$0703&   0155$-$070  & 10,690   &(165) &8.09 (0.06)    &0.66 (0.00)    &12.10          &$-$2.76        &16.2        &   66&5.69($-$6)     &8.76           &                    \\
 01595$-$1109&   0159$-$111  & 11,140   &(174) &8.22 (0.06)    &0.74 (0.00)    &12.18          &$-$2.77        &16.9        &   87&6.30($-$6)     &8.80           &                    \\
02185+2923   &  0218+293     & 17,110   &(292) &8.15 (0.05)    &0.71 (0.00)    &11.27          &$-$1.98        &17.50       &  176&            ---&8.23           &                    \\
\\
02196+2816   &  0219+282A    & 27,170   &(383) &8.09 (0.04)    &0.69 (0.00)    &10.32          &$-$1.12        &18.16       &  369&            ---&7.25           &2                   \\
 02245+3145  &   0224+317    & 14,550   &(330) &7.77 (0.06)    &0.49 (0.00)    &11.00          &$-$2.04        &16.9        &  151&1.42($-$6)     &8.18           &                    \\
 02292+2704  &   0229+270    & 24,270   &(374) &7.88 (0.05)    &0.57 (0.00)    &10.24          &$-$1.19        &15.58       &  116&5.72($-$7)     &7.31           &                    \\
 02295+3529  &   0229+354    & 12,860   &(385) &7.78 (0.07)    &0.49 (0.00)    &11.26          &$-$2.26        &16.90       &  134&1.96($-$6)     &8.36           &                    \\
 02306+3420  &   0230+343    & 14,600   &(242) &7.85 (0.05)    &0.53 (0.00)    &11.11          &$-$2.08        &16.0        &   95&1.63($-$6)     &8.23           &                    \\
\\
 02386+3322  &   0238+333    & 13,390   &(276) &8.23 (0.05)    &0.75 (0.00)    &11.82          &$-$2.45        &15.9        &   65&3.97($-$6)     &8.59           &                    \\
 02464+3239  &   0246+326    & 11,540   &(173) &8.11 (0.05)    &0.67 (0.00)    &11.91          &$-$2.64        &15.8        &   59&4.46($-$6)     &8.68           &                    \\
 02503$-$0238&   0250$-$026  & 14,740   &(241) &7.87 (0.05)    &0.54 (0.00)    &11.12          &$-$2.07        &14.73       &   52&1.65($-$6)     &8.23           &                    \\
 03036$-$0043&   0303$-$007  & 18,640   &(413) &7.97 (0.07)    &0.61 (0.00)    &10.86          &$-$1.72        &16.21       &  117&1.20($-$6)     &7.94           &1                   \\
 03106$-$0719&   0310$-$073  & 17,490   &(323) &7.89 (0.06)    &0.56 (0.00)    &10.84          &$-$1.78        &16.7        &  148&1.17($-$6)     &7.97           &                    \\
03123+0155   &  0312+019     & 41,140   &(854) &8.02 (0.08)    &0.69 (0.00)    & 9.45          &$-$0.34        &17.5        &  406&            ---&6.57           &                    \\
 03184$-$0211&   0318$-$021  & 12,800   &(217) &7.93 (0.05)    &0.57 (0.00)    &11.47          &$-$2.36        &16.01       &   80&2.55($-$6)     &8.46           &                    \\
 03205$-$0005&   0320$-$000  & 13,180   &(499) &7.73 (0.07)    &0.46 (0.00)    &11.13          &$-$2.19        &16.97       &  147&1.67($-$6)     &8.30           &                    \\
03217$-$0240 &  0321$-$026   & 26,890   &(530) &8.46 (0.08)    &0.92 (0.00)    &10.96          &$-$1.39        &17.67       &  219&            ---&7.83           &                    \\
 03290+0053  &   0328+008    & 34,650   &(564) &7.92 (0.07)    &0.62 (0.00)    & 9.55          &$-$0.58        &16.8        &  282&2.61($-$7)     &6.79           &                    \\
\\
 03292+0035  &   0329+005    & 15,500   &(296) &8.00 (0.06)    &0.61 (0.00)    &11.22          &$-$2.06        &16.7        &  124&1.87($-$6)     &8.25           &3                   \\
 03295$-$0108&   0329$-$011  & 17,070   &(306) &7.76 (0.06)    &0.49 (0.00)    &10.70          &$-$1.75        &17.17       &  196&9.88($-$7)     &7.91           &                    \\
 03301$-$0100&   0330$-$009  & 33,800   &(510) &7.91 (0.05)    &0.61 (0.00)    & 9.57          &$-$0.61        &15.86       &  181&2.66($-$7)     &6.82           &                    \\
 03302$-$0143&   0330$-$017  & 30,290   &(460) &7.86 (0.05)    &0.58 (0.00)    & 9.72          &$-$0.78        &17.12       &  302&3.15($-$7)     &6.98           &                    \\
 03351+0245  &   0335+027    & 16,080   &(289) &7.96 (0.05)    &0.59 (0.00)    &11.09          &$-$1.97        &17.4        &  182&1.59($-$6)     &8.17           &                    \\
\\
 03363+0400  &   0336+040    &  8820    &(129) &8.19 (0.06)    &0.71 (0.00)    &12.96          &$-$3.16        &15.9        &   38&1.74($-$5)     &9.04           &                    \\
03383$-$0146 &  0338$-$017   & 20,500   &(374) &8.18 (0.05)    &0.73 (0.00)    &11.00          &$-$1.67        &17.53       &  202&            ---&7.98           &                    \\
03399+0015   &  0339+002A    & 13,680   &(193) &7.97 (0.04)    &0.59 (0.00)    &11.39          &$-$2.26        &17.72       &  184&            ---&8.39           &2                   \\
 03416+0206  &   0341+021    & 22,310   &(341) &7.46 (0.05)    &0.39 (0.00)    & 9.77          &$-$1.09        &15.8        &  160&3.33($-$7)     &7.36           &                    \\
 03439$-$0048&   0343$-$007  & 61,810   &(1358)&7.84 (0.07)    &0.65 (0.00)    & 8.66          &$+$0.53        &14.91       &  177&1.02($-$7)     &6.06           &4                   \\
\\
 03442+0719  &   0344+073    & 10,930   &(164) &7.84 (0.06)    &0.51 (0.00)    &11.67          &$-$2.58        &16.10       &   77&3.29($-$6)     &8.59           &                    \\
 03459+0037  &   0345+006    & 86,850   &(3544)&7.08 (0.12)    &0.51 (0.00)    & 6.49          &$+$1.77        &16.00       &  799&1.63($-$8)     &4.70           &4, 5                \\
 03463$-$0108&   0346$-$011  & 41,260   &(765) &9.11 (0.07)    &1.26 (0.00)    &11.50          &$-$1.15        &14.06       &   32&2.65($-$6)     &7.93           &                    \\
 03521+0150  &   0352+018    & 22,350   &(353) &7.92 (0.05)    &0.59 (0.00)    &10.46          &$-$1.36        &15.2        &   88&7.41($-$7)     &7.52           &                    \\
 03520+0500  &   0352+049    & 36,270   &(592) &8.77 (0.06)    &1.10 (0.00)    &10.97          &$-$1.10        &16.2        &  111&1.37($-$6)     &7.78           &                    \\
\\
 03520+0515  &   0352+052    & 10,280   &(151) &8.10 (0.06)    &0.66 (0.00)    &12.25          &$-$2.84        &15.9        &   53&6.90($-$6)     &8.81           &                    \\
 03522+0737  &   0352+076    & 15,230   &(278) &7.74 (0.06)    &0.48 (0.00)    &10.89          &$-$1.94        &16.76       &  149&1.25($-$6)     &8.09           &                    \\
 03526+0939  &   0352+096    & 14,030   &(341) &8.19 (0.05)    &0.73 (0.00)    &11.68          &$-$2.35        &14.56       &   37&3.33($-$6)     &8.51           &                    \\
 03561+0807  &   0356+081    & 43,780   &(840) &8.05 (0.07)    &0.70 (0.00)    & 9.42          &$-$0.24        &16.7        &  286&2.26($-$7)     &6.50           &                    \\
 04100+1144  &   0410+117    & 20,630   &(326) &7.95 (0.05)    &0.60 (0.00)    &10.65          &$-$1.53        &14.03       &   47&9.30($-$7)     &7.72           &                    \\
\\
 04190+1522  &   0418+153    & 13,490   &(346) &7.99 (0.05)    &0.60 (0.00)    &11.45          &$-$2.30        &16.62       &  107&2.49($-$6)     &8.43           &                    \\
 04211+1614  &   0421+162    & 19,140   &(299) &8.05 (0.05)    &0.65 (0.00)    &10.92          &$-$1.72        &14.27       &   46&1.29($-$6)     &7.97           &                    \\
 04234+1222  &   0423+123    & 21,360   &(497) &7.94 (0.07)    &0.60 (0.00)    &10.57          &$-$1.46        &16.9        &  184&8.45($-$7)     &7.64           &                    \\
04239+1406   &  0423+140     & 12,820   &(454) &8.89 (0.08)    &1.15 (0.00)    &13.05          &$-$3.01        &17.56       &   79&            ---&9.17           &                    \\
 04258+1652  &   0425+168    & 24,200   &(371) &8.07 (0.05)    &0.67 (0.00)    &10.54          &$-$1.32        &13.92       &   47&8.15($-$7)     &7.51           &                    \\
 04262+1038  &   0426+106    & 10,380   &(154) &8.66 (0.05)    &1.02 (0.00)    &13.16          &$-$3.19        &16.3        &   42&2.27($-$5)     &9.26           &                    \\
 04295+1739  &   0429+176    & 17,540   &(481) &7.98 (0.08)    &0.61 (0.00)    &10.97          &$-$1.83        &13.93       &   39&1.37($-$6)     &8.05           &1                   \\
 04304+1339  &   0430+136    & 35,720   &(837) &8.11 (0.12)    &0.72 (0.00)    & 9.80          &$-$0.65        &16.50       &  218&3.44($-$7)     &6.81           &1                   \\
 04310+1236  &   0431+126    & 20,810   &(327) &8.09 (0.05)    &0.68 (0.00)    &10.84          &$-$1.59        &14.24       &   47&1.17($-$6)     &7.86           &                    \\
 04370+1514  &   0437+152    & 18,660   &(322) &7.39 (0.05)    &0.35 (0.00)    &10.00          &$-$1.36        &15.83       &  146&4.33($-$7)     &7.50           &                    \\
\\
 04383+1054  &   0438+108    & 26,820   &(394) &8.08 (0.05)    &0.68 (0.00)    &10.33          &$-$1.14        &13.86       &   50&6.35($-$7)     &7.26           &                    \\
06548+3225   &  0654+324     & 22,020   &(386) &8.09 (0.05)    &0.68 (0.00)    &10.74          &$-$1.50        &18.1        &  295&            ---&7.75           &                    \\
 07069+2929  &   0706+294    & 14,040   &(253) &7.84 (0.05)    &0.52 (0.00)    &11.17          &$-$2.14        &15.45       &   71&1.76($-$6)     &8.28           &                    \\
 07170+3653  &   0717+368    & 23,870   &(470) &7.97 (0.06)    &0.62 (0.00)    &10.41          &$-$1.28        &16.8        &  190&6.99($-$7)     &7.42           &                    \\
 07540+4015  &   0754+402    & 18,830   &(481) &7.89 (0.08)    &0.56 (0.00)    &10.72          &$-$1.65        &16.7        &  157&1.01($-$6)     &7.84           &                    \\
\\
 08016+4206  &   0801+421    & 13,730   &(560) &8.14 (0.07)    &0.69 (0.00)    &11.63          &$-$2.35        &17.06       &  121&3.12($-$6)     &8.50           &                    \\
 08026+4118  &   0802+413    & 51,890   &(977) &7.60 (0.06)    &0.53 (0.00)    & 8.43          &$+$0.38        &14.5        &  163&8.14($-$8)     &6.27           &                    \\
 08039+4003  &   0803+400    & 12,450   &(247) &8.07 (0.06)    &0.65 (0.00)    &11.71          &$-$2.48        &17.40       &  137&3.46($-$6)     &8.57           &                    \\
 08084+4221  &   0808+423    & 14,900   &(446) &8.90 (0.05)    &1.15 (0.00)    &12.81          &$-$2.75        &16.76       &   61&1.43($-$5)     &9.00           &                    \\
 08100+3915  &   0810+392    & 22,540   &(453) &7.91 (0.06)    &0.58 (0.00)    &10.42          &$-$1.34        &16.85       &  193&7.07($-$7)     &7.48           &                    \\
\\
 08157+3739  &   0815+376    & 21,690   &(365) &7.95 (0.05)    &0.60 (0.00)    &10.55          &$-$1.43        &16.7        &  170&8.25($-$7)     &7.61           &                    \\
 08157+3946  &   0815+397    & 37,820   &(727) &7.85 (0.08)    &0.59 (0.00)    & 9.29          &$-$0.37        &17.40       &  419&1.96($-$7)     &6.65           &                    \\
 08165+3741  &   0816+376    & 11,000   &(627) &8.00 (0.06)    &0.60 (0.00)    &11.88          &$-$2.66        &15.6        &   55&4.29($-$6)     &8.67           &6                   \\
 08167+3844  &   0816+387    &  7630    &(112) &7.98 (0.07)    &0.58 (0.00)    &13.22          &$-$3.29        &16.55       &   46&2.46($-$5)     &9.07           &                    \\
 08172+3838  &   0817+386    & 25,330   &(388) &8.03 (0.05)    &0.65 (0.00)    &10.38          &$-$1.21        &15.69       &  115&6.74($-$7)     &7.35           &                    \\
\\
 08268+4150  &   0826+418    & 10,270   &(158) &8.12 (0.07)    &0.67 (0.00)    &12.28          &$-$2.85        &16.8        &   80&7.17($-$6)     &8.82           &                    \\
08275+3252   &  0827+328     &  7490    &(115) &8.63 (0.08)    &1.00 (0.00)    &14.32          &$-$3.74        &15.73       &   19&            ---&9.56           &                    \\
 08273+4101  &   0827+410    & 15,210   &(254) &7.77 (0.05)    &0.49 (0.00)    &10.93          &$-$1.96        &15.92       &   99&1.31($-$6)     &8.12           &                    \\
 08308+3710  &   0830+371    &  9180    &(133) &8.26 (0.06)    &0.76 (0.00)    &12.92          &$-$3.13        &16.01       &   41&1.65($-$5)     &9.04           &                    \\
 08317+4117  &   0831+412    & 31,300   &(510) &8.43 (0.07)    &0.90 (0.00)    &10.57          &$-$1.10        &17.11       &  202&8.45($-$7)     &7.54           &                    \\
\\
08354+3639   &  0835+366     & 15,690   &(284) &7.91 (0.05)    &0.56 (0.00)    &11.07          &$-$1.98        &17.87       &  229&            ---&8.17           &                    \\
 08368+4026  &   0836+404    & 11,870   &(180) &8.10 (0.05)    &0.67 (0.00)    &11.84          &$-$2.59        &15.55       &   55&4.08($-$6)     &8.65           &                    \\
08371+3754   &  0837+378     & 14,340   &(440) &8.17 (0.06)    &0.71 (0.00)    &11.61          &$-$2.30        &17.50       &  150&            ---&8.47           &                    \\
 08378+3934  &   0837+395    & 20,420   &(485) &7.89 (0.07)    &0.56 (0.00)    &10.57          &$-$1.50        &16.54       &  156&8.45($-$7)     &7.67           &                    \\
 08381+3737  &   0838+376    & 19,040   &(328) &7.88 (0.05)    &0.56 (0.00)    &10.68          &$-$1.62        &16.99       &  182&9.65($-$7)     &7.80           &                    \\
 08397+3435  &   0839+345    & 17,550   &(286) &8.06 (0.05)    &0.65 (0.00)    &11.08          &$-$1.87        &16.2        &  105&1.57($-$6)     &8.11           &                    \\
 08391+3800  &   0839+379    & 19,020   &(376) &8.06 (0.06)    &0.65 (0.00)    &10.94          &$-$1.73        &16.32       &  119&1.32($-$6)     &7.99           &                    \\
 08411+3340  &   0841+336    & 23,990   &(461) &8.12 (0.06)    &0.70 (0.00)    &10.63          &$-$1.36        &16.7        &  163&9.08($-$7)     &7.60           &                    \\
 08417+4112  &   0841+411    & 16,070   &(336) &7.92 (0.06)    &0.57 (0.00)    &11.04          &$-$1.95        &17.40       &  187&1.50($-$6)     &8.14           &                    \\
 08422+3813  &   0842+382    &  7990    &(116) &8.06 (0.06)    &0.63 (0.00)    &13.15          &$-$3.26        &16.03       &   37&2.24($-$5)     &9.07           &4                   \\
\\
 08460+3441  &   0846+346    &  7600    &(110) &8.07 (0.06)    &0.63 (0.00)    &13.36          &$-$3.35        &15.72       &   29&2.95($-$5)     &9.13           &                    \\
08473+3838   &  0847+386     & 17,280   &(286) &8.30 (0.05)    &0.80 (0.00)    &11.48          &$-$2.05        &17.67       &  172&            ---&8.32           &                    \\
 08504+4155  &   0850+419    & 18,200   &(321) &7.87 (0.05)    &0.55 (0.00)    &10.75          &$-$1.70        &16.9        &  169&1.05($-$6)     &7.89           &                    \\
 08543+4028  &   0854+404    & 22,270   &(341) &7.93 (0.05)    &0.59 (0.00)    &10.47          &$-$1.37        &14.90       &   76&7.50($-$7)     &7.53           &                    \\
 08587+3619  &   0858+363    & 11,970   &(178) &8.17 (0.05)    &0.71 (0.00)    &11.93          &$-$2.62        &14.55       &   33&4.57($-$6)     &8.68           &                    \\
\\
 09029+4153  &   0902+418    & 44,040   &(1029)&8.05 (0.09)    &0.71 (0.00)    & 9.42          &$-$0.23        &17.2        &  360&2.26($-$7)     &6.49           &                    \\
09272+3930   &  0927+394     & 23,940   &(486) &8.32 (0.07)    &0.82 (0.00)    &10.95          &$-$1.49        &17.7        &  224&            ---&7.87           &                    \\
09288+3959   &  0928+399     & 25,300   &(503) &8.04 (0.07)    &0.66 (0.00)    &10.40          &$-$1.22        &17.53       &  267&            ---&7.36           &                    \\
 09443+4229  &   0944+424    & 23,680   &(368) &8.02 (0.05)    &0.64 (0.00)    &10.49          &$-$1.32        &16.43       &  154&7.68($-$7)     &7.49           &                    \\
 09479+3234  &   0947+325    & 22,200   &(345) &8.31 (0.05)    &0.81 (0.00)    &11.07          &$-$1.62        &15.38       &   72&1.55($-$6)     &7.98           &                    \\
\\
 09538+3405  &   0953+340    & 16,640   &(395) &8.01 (0.07)    &0.62 (0.00)    &11.11          &$-$1.94        &16.76       &  134&1.63($-$6)     &8.16           &                    \\
 09583+3520  &   0958+353    & 40,940   &(897) &7.93 (0.09)    &0.64 (0.00)    & 9.31          &$-$0.28        &16.9        &  330&2.00($-$7)     &6.56           &                    \\
 10010+3318  &   1001+333    &  9760    &(143) &8.21 (0.06)    &0.73 (0.00)    &12.61          &$-$3.00        &16.3        &   54&1.10($-$5)     &8.94           &                    \\
 10013+3614  &   1001+362    &  9450    &(142) &7.33 (0.09)    &0.30 (0.00)    &11.47          &$-$2.57        &16.8        &  116&2.55($-$6)     &8.56           &                    \\
10063+3522   &  1006+353     & 19,920   &(461) &7.85 (0.07)    &0.55 (0.00)    &10.56          &$-$1.53        &17.83       &  283&            ---&7.68           &                    \\
\\
 10090+3712  &   1008+372    & 15,150   &(272) &7.91 (0.05)    &0.56 (0.00)    &11.12          &$-$2.04        &16.8        &  136&1.65($-$6)     &8.21           &                    \\
 10081+3817  &   1008+382    & 13,160   &(356) &7.87 (0.06)    &0.54 (0.00)    &11.33          &$-$2.27        &16.29       &   98&2.14($-$6)     &8.38           &                    \\
 10115+3332  &   1011+335    & 19,690   &(404) &8.11 (0.06)    &0.69 (0.00)    &10.97          &$-$1.71        &17.4        &  193&1.37($-$6)     &7.98           &                    \\
 11230+4240  &   1123+426    & 10,400   &(156) &8.18 (0.06)    &0.71 (0.00)    &12.33          &$-$2.87        &17.0        &   86&7.65($-$6)     &8.85           &                    \\
 11265+3825  &   1126+384    & 25,060   &(388) &7.96 (0.05)    &0.61 (0.00)    &10.30          &$-$1.18        &14.89       &   82&6.13($-$7)     &7.30           &                    \\
\\
 11370+4222  &   1137+423    & 11,840   &(178) &8.15 (0.05)    &0.70 (0.00)    &11.92          &$-$2.62        &16.56       &   84&4.52($-$6)     &8.68           &                    \\
 11390+4225  &   1139+424    & 27,250   &(425) &7.90 (0.05)    &0.59 (0.00)    &10.02          &$-$1.00        &16.20       &  172&4.43($-$7)     &7.10           &                    \\
11472+3858   &  1147+389     & 17,220   &(311) &7.89 (0.05)    &0.56 (0.00)    &10.87          &$-$1.81        &17.46       &  207&            ---&8.00           &                    \\
 11491+4104  &   1149+410    & 14,070   &(271) &7.84 (0.05)    &0.52 (0.00)    &11.17          &$-$2.14        &16.08       &   96&1.76($-$6)     &8.28           &                    \\
12279+3044   &  1227+307     & 13,330   &(544) &7.93 (0.07)    &0.57 (0.00)    &11.39          &$-$2.28        &17.7        &  182&            ---&8.40           &                    \\
 12353+2925  &   1235+294    & 18,850   &(357) &7.90 (0.06)    &0.56 (0.00)    &10.72          &$-$1.65        &17.29       &  205&1.01($-$6)     &7.84           &                    \\
 12399+2744  &   1239+277    & 14,740   &(256) &7.84 (0.05)    &0.52 (0.00)    &11.08          &$-$2.06        &15.7        &   83&1.57($-$6)     &8.21           &                    \\
 12420+2938  &   1241+296    & 17,870   &(354) &7.81 (0.06)    &0.52 (0.00)    &10.70          &$-$1.70        &16.0        &  114&9.88($-$7)     &7.87           &                    \\
12436+3011   &  1243+301     & 14,140   &(369) &7.75 (0.07)    &0.48 (0.00)    &11.03          &$-$2.08        &17.60       &  206&            ---&8.21           &                    \\
 12474+3105  &   1247+310    & 11,940   &(216) &8.34 (0.06)    &0.82 (0.00)    &12.19          &$-$2.72        &17.2        &  100&6.38($-$6)     &8.80           &                    \\
\\
 12492+2937  &   1249+296    & 11,770   &(211) &8.25 (0.06)    &0.76 (0.00)    &12.09          &$-$2.69        &16.8        &   87&5.61($-$6)     &8.76           &                    \\
 12574+2750  &   1257+278    &  8710    &(126) &8.36 (0.06)    &0.82 (0.00)    &13.28          &$-$3.29        &15.39       &   26&2.66($-$5)     &9.19           &                    \\
12587+2942   &  1258+297     & 14,830   &(306) &7.91 (0.06)    &0.56 (0.00)    &11.16          &$-$2.08        &18.3        &  267&            ---&8.25           &                    \\
 13088+3139  &   1308+316    & 13,260   &(444) &7.93 (0.06)    &0.57 (0.00)    &11.40          &$-$2.29        &16.79       &  119&2.34($-$6)     &8.41           &                    \\
 14083+3223  &   1408+323    & 18,160   &(279) &7.91 (0.05)    &0.57 (0.00)    &10.81          &$-$1.73        &14.11       &   45&1.13($-$6)     &7.93           &                    \\
\\
 14134+2311  &   1413+231    & 23,290   &(412) &7.74 (0.05)    &0.50 (0.00)    &10.10          &$-$1.18        &16.9        &  228&4.86($-$7)     &7.33           &                    \\
 14138+2408  &   1413+241    & 16,450   &(265) &8.05 (0.05)    &0.65 (0.00)    &11.19          &$-$1.98        &17.0        &  145&1.80($-$6)     &8.20           &                    \\
14197+2514   &  1419+252A    & 10,160   &(143) &8.00 (0.04)    &0.60 (0.00)    &12.14          &$-$2.80        &18.10       &  155&            ---&8.76           &2                   \\
 14205+2251  &   1420+228    & 17,300   &(319) &7.83 (0.06)    &0.52 (0.00)    &10.78          &$-$1.76        &16.8        &  160&1.09($-$6)     &7.94           &                    \\
 14227+3340  &   1422+336    & 13,740   &(430) &8.15 (0.06)    &0.70 (0.00)    &11.66          &$-$2.36        &17.1        &  122&3.24($-$6)     &8.51           &                    \\
\\
 14287+3724  &   1428+373    & 14,010   &(221) &7.36 (0.05)    &0.32 (0.00)    &10.50          &$-$1.87        &15.6        &  104&7.77($-$7)     &8.07           &                    \\
 14299+3720  &   1429+373    & 34,390   &(507) &8.11 (0.05)    &0.72 (0.00)    & 9.86          &$-$0.71        &15.27       &  120&3.69($-$7)     &6.88           &                    \\
 14310+2542  &   1431+257    & 22,610   &(399) &7.21 (0.05)    &0.32 (0.00)    & 9.32          &$-$0.89        &17.0        &  342&2.03($-$7)     &7.29           &                    \\
 15502+1819  &   1550+183    & 14,260   &(271) &8.25 (0.05)    &0.77 (0.00)    &11.75          &$-$2.36        &14.9        &   42&3.64($-$6)     &8.53           &                    \\
 16055+1745  &   1605+177    & 13,830   &(267) &7.68 (0.05)    &0.44 (0.00)    &10.97          &$-$2.07        &16.7        &  140&1.37($-$6)     &8.20           &                    \\
\\
 16069+1810  &   1606+181    & 22,150   &(373) &7.98 (0.05)    &0.62 (0.00)    &10.56          &$-$1.42        &17.1        &  202&8.35($-$7)     &7.60           &                    \\
 16075+2031  &   1607+205    & 11,150   &(173) &7.82 (0.06)    &0.50 (0.00)    &11.58          &$-$2.53        &17.4        &  145&2.93($-$6)     &8.56           &                    \\
 16106+3820  &   1610+383    & 14,450   &(278) &7.83 (0.05)    &0.52 (0.00)    &11.10          &$-$2.08        &16.4        &  114&1.61($-$6)     &8.23           &                    \\
 16195+4125  &   1619+414    & 14,090   &(457) &7.93 (0.06)    &0.57 (0.00)    &11.28          &$-$2.18        &16.8        &  126&2.01($-$6)     &8.33           &1                   \\
 16268+4055  &   1626+409    & 21,400   &(422) &7.98 (0.06)    &0.62 (0.00)    &10.63          &$-$1.48        &16.7        &  163&9.08($-$7)     &7.68           &                    \\
\\
 16288+3904  &   1628+390    & 19,040   &(340) &7.88 (0.05)    &0.56 (0.00)    &10.68          &$-$1.62        &16.8        &  167&9.65($-$7)     &7.80           &                    \\
 16319+3937  &   1631+396    & 17,380   &(281) &7.63 (0.05)    &0.43 (0.00)    &10.49          &$-$1.64        &14.1        &   52&7.68($-$7)     &7.80           &                    \\
 16366+3506  &   1636+351    & 36,950   &(561) &8.04 (0.05)    &0.69 (0.00)    & 9.63          &$-$0.54        &14.9        &  113&2.85($-$7)     &6.73           &                    \\
 16376+3331  &   1637+335    & 10,260   &(147) &8.20 (0.05)    &0.72 (0.00)    &12.41          &$-$2.90        &14.8        &   30&8.49($-$6)     &8.88           &                    \\
 16476+3733  &   1647+375    & 21,860   &(343) &7.92 (0.05)    &0.59 (0.00)    &10.50          &$-$1.40        &14.98       &   78&7.77($-$7)     &7.57           &                    \\
 16484+3706  &   1648+371    & 42,240   &(904) &7.63 (0.08)    &0.51 (0.00)    & 8.75          &$-$0.02        &15.91       &  270&1.12($-$7)     &6.53           &                    \\
 18004+6836  &   1800+685    & 44,110   &(783) &7.90 (0.06)    &0.63 (0.00)    & 9.16          &$-$0.13        &14.60       &  122&1.71($-$7)     &6.47           &4                   \\
 18284+6650  &   1828+668    & 10,800   &(163) &8.20 (0.06)    &0.73 (0.00)    &12.24          &$-$2.81        &16.65       &   76&6.81($-$6)     &8.82           &                    \\
 18332+6429  &   1833+644    & 46,300   &(1764)&8.12 (0.14)    &0.75 (0.00)    & 9.48          &$-$0.19        &16.9        &  304&2.41($-$7)     &6.42           &1                   \\
 18453+6819  &   1845+683    & 36,810   &(574) &8.31 (0.05)    &0.84 (0.00)    &10.08          &$-$0.73        &15.00       &   96&4.75($-$7)     &6.88           &4                   \\
\\
 21168+7338  &   2116+736    & 52,810   &(1182)&7.66 (0.08)    &0.56 (0.00)    & 8.52          &$+$0.37        &14.87       &  186&8.88($-$8)     &6.27           &                    \\
 21267+7326  &   2126+734    & 15,290   &(223) &7.84 (0.04)    &0.53 (0.00)    &11.02          &$-$1.99        &12.78       &   22&1.46($-$6)     &8.16           &                    \\
 22543+1237  &   2254+126    & 11,640   &(173) &8.05 (0.05)    &0.64 (0.00)    &11.81          &$-$2.59        &16.0        &   68&3.92($-$6)     &8.64           &                    \\
 22570+1349  &   2257+138    & 27,300   &(399) &8.36 (0.05)    &0.85 (0.00)    &10.74          &$-$1.29        &16.65       &  151&1.04($-$6)     &7.66           &                    \\
 22573+1613  &   2257+162    & 24,840   &(386) &7.49 (0.05)    &0.41 (0.00)    & 9.60          &$-$0.91        &16.14       &  203&2.75($-$7)     &7.23           &                    \\
\\
 23060+1303  &   2306+124    & 20,340   &(322) &8.06 (0.05)    &0.66 (0.00)    &10.84          &$-$1.62        &15.23       &   75&1.17($-$6)     &7.88           &                    \\
 23061+1229  &   2306+130    & 13,250   &(283) &7.92 (0.05)    &0.56 (0.00)    &11.38          &$-$2.29        &15.1        &   55&2.28($-$6)     &8.40           &                    \\
 23083+1642  &   2308+167    & 18,000   &(322) &7.95 (0.05)    &0.59 (0.00)    &10.88          &$-$1.76        &17.2        &  184&1.23($-$6)     &7.97           &                    \\
 23098+1031  &   2309+105    & 54,150   &(951) &8.00 (0.06)    &0.70 (0.00)    & 9.09          &$+$0.17        &13.09       &   63&1.59($-$7)     &6.20           &                    \\
23128+1157   &  2312+119     & 18,100   &(306) &7.70 (0.05)    &0.46 (0.00)    &10.51          &$-$1.61        &17.83       &  291&            ---&7.76           &                    \\
\\
 23149+1408  &   2314+141    & 17,840   &(304) &7.78 (0.05)    &0.50 (0.00)    &10.65          &$-$1.68        &16.9        &  177&9.30($-$7)     &7.85           &                    \\
 23162+1220  &   2316+123    & 11,000   &(383) &8.00 (0.06)    &0.60 (0.00)    &11.88          &$-$2.66        &15.38       &   50&4.29($-$6)     &8.67           &7                   \\
 23176+2650  &   2317+268    & 31,480   &(491) &7.68 (0.06)    &0.50 (0.00)    & 9.34          &$-$0.59        &16.3        &  246&2.07($-$7)     &6.91           &                    \\
 23189+0901  &   2318+090    & 28,990   &(475) &7.97 (0.06)    &0.63 (0.00)    &10.00          &$-$0.93        &16.27       &  179&4.33($-$7)     &7.02           &                    \\
 23180+1242  &   2318+126    & 13,490   &(334) &7.92 (0.05)    &0.56 (0.00)    &11.35          &$-$2.25        &16.27       &   96&2.20($-$6)     &8.38           &                    \\
\\
 23220+0921  &   2322+093    & 14,060   &(304) &7.81 (0.06)    &0.51 (0.00)    &11.12          &$-$2.12        &16.7        &  130&1.65($-$6)     &8.26           &                    \\
 23223+0953  &   2322+098    & 20,090   &(341) &7.88 (0.05)    &0.56 (0.00)    &10.59          &$-$1.53        &17.2        &  210&8.66($-$7)     &7.69           &                    \\
23235+2536   &  2323+256     &  5930    &(279) &7.85 (0.66)    &0.50 (0.00)    &14.11          &$-$3.67        &17.02       &   38&            ---&9.27           &                    \\
 23282+1046  &   2328+107    & 22,180   &(363) &7.85 (0.05)    &0.55 (0.00)    &10.37          &$-$1.34        &15.53       &  107&6.66($-$7)     &7.48           &                    \\
 23296+2642  &   2329+267    &  9400    &(274) &8.02 (0.28)    &0.61 (0.00)    &12.46          &$-$2.95        &15.4        &   38&9.06($-$6)     &8.86           &8                   \\
\\
\enddata
\tablecomments{
(1) DA+dM; 
(2) Unresolved double degenerate; 
(3) Magnetic, $B=12$ MG \citep{jordan93}, $\Te$              from \citealt{schmidt03}; 
(4) Not in \citealt{darling94}; (5) DAO star; 
(6) Magnetic, $B=9$ MG (\citet{angel74}, $\Te$ from \citealt{jordan01}; 
(7) Magnetic, $B=29$ MG and $\Te$ from \citet{liebert85}; 
(8) Magnetic, B (2.3 MG), $\Te$ and $\logg$ from             \citealt{bergeron01}. }
\end{deluxetable}

\clearpage
\begin{deluxetable}{llrrlccrclccclc}
\tabletypesize{\scriptsize}
\tablecolumns{12}
\tablewidth{0pt}
\rotate
\tablecaption{Atmospheric Parameters of DB Stars from the KUV Sample}
\tablehead{
\colhead{KUV} &
\colhead{WD} &
\colhead{$\Te$} &
\colhead{(K)} &
\colhead{log $g$} &
\colhead{log $\rm{H/He}$} &
\colhead{$M/M_{\odot}$} &
\colhead{$M_V$} &
\colhead{log $L/L_{\odot}$} &
\colhead{$V$} &
\colhead{$D ({\rm pc})$} &
\colhead{$1/v_{\rm max}$} &
\colhead{log $\tau$} &
\colhead{Notes}}
\startdata
 00312$-$1837&   0031$-$186  & 15,270   &(249) &8.43 (0.09)    &---            &0.86 (0.00)    &11.93          &$-$2.36        &16.66       &   88&4.64($-$6)     &8.59           &                    \\
 01254$-$2340&   0125$-$236  & 16,500   &(261) &8.29 (0.08)    &  $-$4.80      &0.77 (0.00)    &11.53          &$-$2.14        &15.38       &   58&2.79($-$6)     &8.39           &                    \\
02196+2816   &  0219+282B    & 36,340   &(512) &8.09 (0.04)    &---            &0.68 (0.00)    &10.11          &$-$0.62        &17.90       &  361&            ---&6.96           &1                   \\
 02499+3442  &   0249+346    & 13,440   &(350) &9.02 (0.18)    &  $-$4.64      &1.20 (0.00)    &13.35          &$-$3.04        &16.40       &   40&2.96($-$5)     &9.14           &                    \\
 02498$-$0515&   0249$-$052  & 17,400   &(265) &8.05 (0.06)    &---            &0.62 (0.00)    &11.03          &$-$1.90        &16.60       &  129&1.50($-$6)     &8.15           &                    \\
\\
 03003$-$0120&   0300$-$013  & 14,280   &(298) &7.91 (0.18)    &---            &0.54 (0.00)    &11.30          &$-$2.17        &15.56       &   71&2.09($-$6)     &8.32           &                    \\
03399+0015   &  0339+002B    & 13,730   &(194) &8.00 (0.04)    &---            &0.59 (0.00)    &11.57          &$-$2.29        &17.82       &  177&            ---&8.44           &1                   \\
 03493+0131  &   0349+015    & 25,300   &(700) &7.98 (0.05)    &---            &0.60 (0.00)    &10.42          &$-$1.20        &17.2        &  226&7.16($-$7)     &7.37           &                    \\
 04376+1353  &   0437+138    & 15,330   &(239) &8.36 (0.08)    &  $-$4.55      &0.82 (0.00)    &11.82          &$-$2.32        &14.92       &   41&4.03($-$6)     &8.54           &                    \\
 05034+1445  &   0503+147    & 15,460   &(229) &8.03 (0.07)    &  $-$4.86      &0.61 (0.00)    &11.28          &$-$2.10        &13.08       &   22&2.04($-$6)     &8.31           &                    \\
\\
 05134+2605  &   0513+260    & 26,170   &(549) &8.24 (0.05)    &---            &0.75 (0.00)    &10.77          &$-$1.30        &16.70       &  153&1.09($-$6)     &7.57           &                    \\
 08381+3737  &   0838+375    & 13,640   &(390) &8.24 (0.25)    &---            &0.73 (0.00)    &11.88          &$-$2.44        &17.         &  105&4.35($-$6)     &8.59           &                    \\
10064+4120   &  1006+413     & 15,170   &(358) &8.86 (0.18)    &  $-$5.09      &1.12 (0.00)    &12.75          &$-$2.69        &17.83       &  103&            ---&8.95           &                    \\
 10098+4138  &   1009+416    & 16,430   &(258) &8.69 (0.08)    &  $-$4.83      &1.02 (0.00)    &12.24          &$-$2.42        &16.33       &   65&6.91($-$6)     &8.68           &                    \\
 11489+4052  &   1148+408    & 17,230   &(321) &8.39 (0.11)    &  $-$4.48      &0.84 (0.00)    &11.60          &$-$2.13        &17.33       &  139&3.05($-$6)     &8.42           &                    \\
\\
 14156+2325  &   1415+234    & 17,590   &(270) &8.17 (0.06)    &---            &0.70 (0.00)    &11.20          &$-$1.95        &16.8        &  132&1.85($-$6)     &8.22           &                    \\
 14161+2255  &   1416+229    & 17,590   &(276) &8.18 (0.07)    &---            &0.70 (0.00)    &11.21          &$-$1.96        &16.9        &  119&1.87($-$6)     &8.22           &1                   \\
 14197+2514  &   1419+252B   & 14,520   &(205) &8.00 (0.04)    &---            &0.59 (0.00)    &11.44          &$-$2.19        &17.34       &  151&2.49($-$6)     &8.37           &1                   \\
 14200+3509  &   1419+351    & 12,770   &(547) &8.85 (0.35)    &  $-$4.57      &1.12 (0.00)    &13.12          &$-$2.99        &16.0        &   37&2.18($-$5)     &9.14           &                    \\
15519+1730   &  1551+175     & 15,440   &(268) &7.89 (0.13)    &  $-$4.34      &0.53 (0.00)    &11.09          &$-$2.02        &17.5        &  191&            ---&8.19           &                    \\
\\
 15571+1913  &   1557+192    & 19,380   &(375) &8.17 (0.06)    &  $-$3.78      &0.70 (0.00)    &11.00          &$-$1.78        &15.4        &   75&1.44($-$6)     &8.07           &                    \\
 16454+3234  &   1645+325    & 25,170   &(472) &7.94 (0.05)    &---            &0.58 (0.00)    &10.37          &$-$1.18        &14.0        &   53&6.75($-$7)     &7.35           &                    \\
 23103+1736  &   2310+175    & 15,170   &(270) &8.29 (0.15)    &  $-$6.49      &0.77 (0.00)    &11.73          &$-$2.29        &15.88       &   67&3.59($-$6)     &8.50           &                    \\
\enddata
\tablecomments{
(1) Unresolved double degenerate. }
\end{deluxetable}

\clearpage

\figcaption[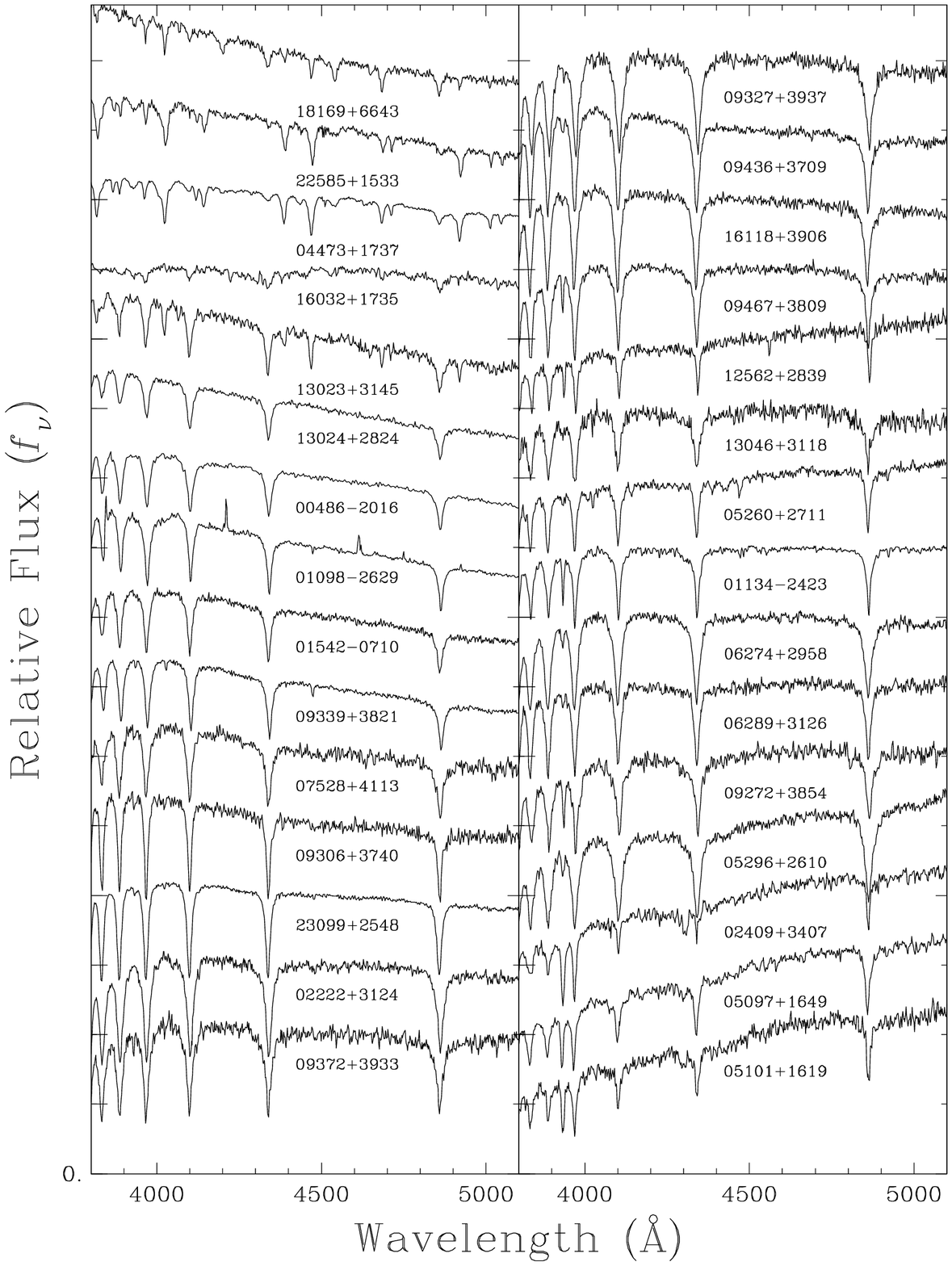]{Optical spectra of hot subdwarfs or main sequence 
stars misclassified as DA or DB stars in the Kiso survey \citep[][and
references therein]{darling94}. The objects are approximately ordered
as a function of their slope. KUV 16032+1735 (left panel, fourth
object from the top) is a hot sdO star.
\label{fg:f1}}

\figcaption[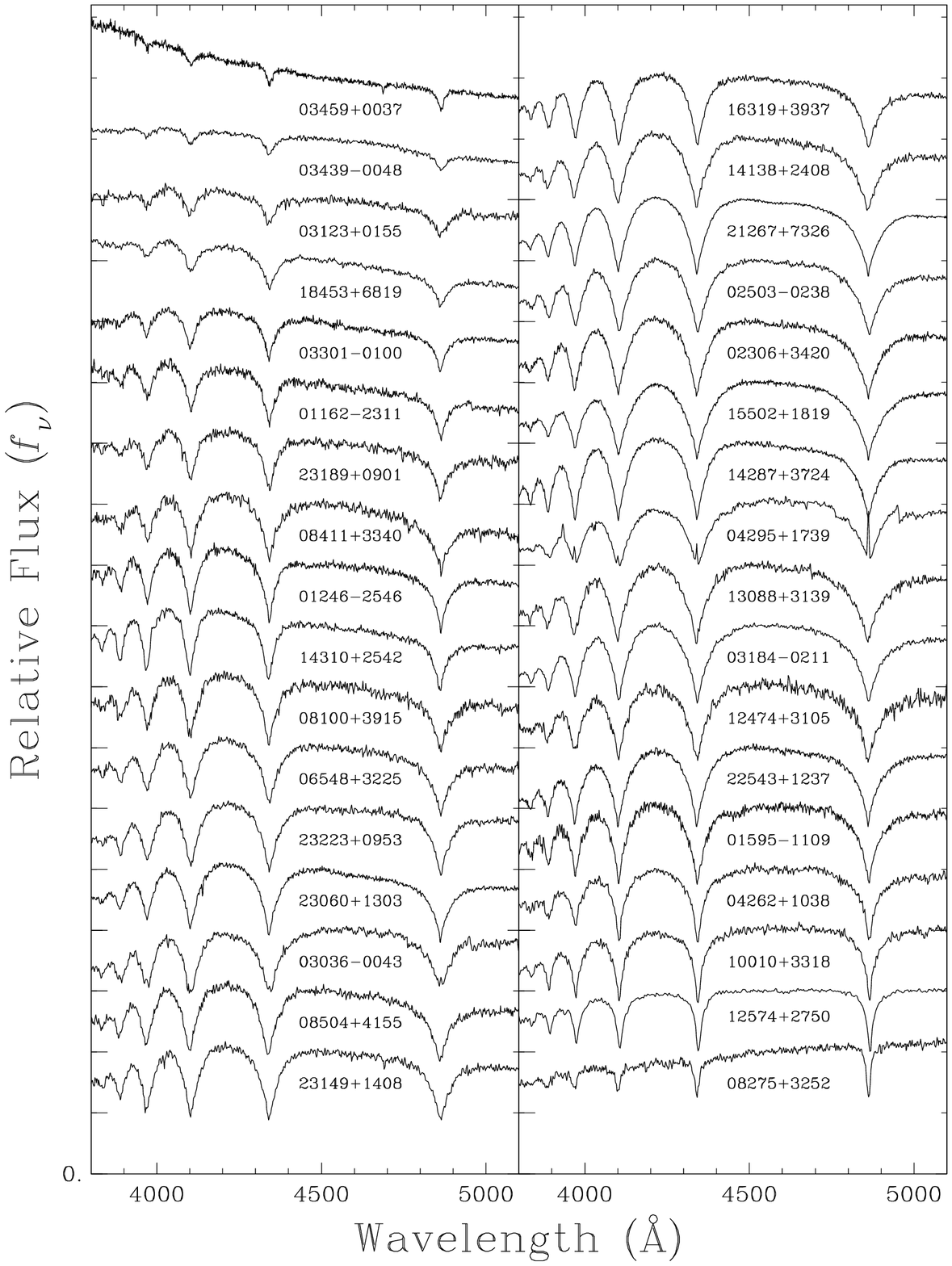]
{Optical spectra for a subsample of DA white dwarfs from the Kiso
survey.  The spectra are normalized at 4500 \AA\ and are shifted
vertically for clarity; the various zero points are indicated by long
tick marks. The effective temperature decreases from upper left to
bottom right. The hottest object in this sample is the DAO star KUV
03459+0037 displayed at the top of the left panel.
\label{fg:f2}}

\figcaption[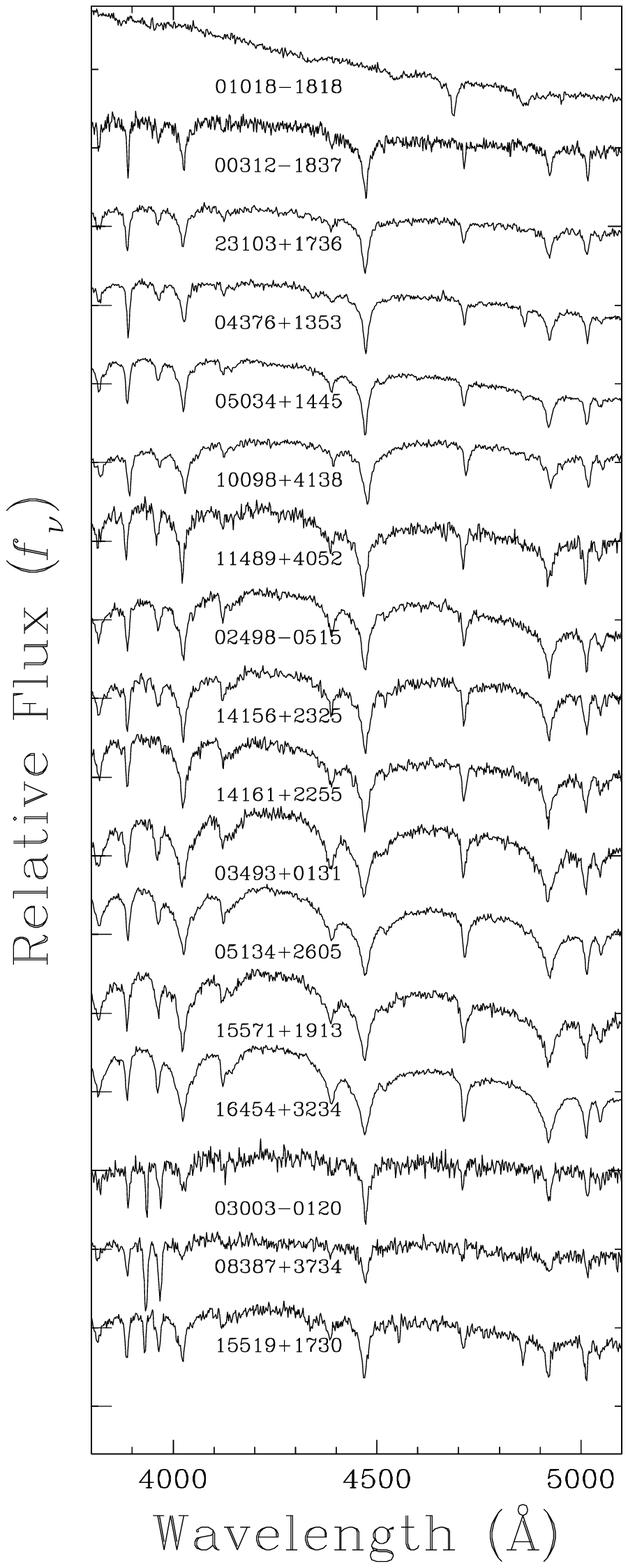]
{Same as Figure \ref{fg:f2} but for a subsample of
DO/DB(Z)/DBA(Z) white dwarfs from the Kiso survey.  The hottest object in
this sample is the DO star KUV 01018$-$1818 shown at the top.
\label{fg:f3}}

\figcaption[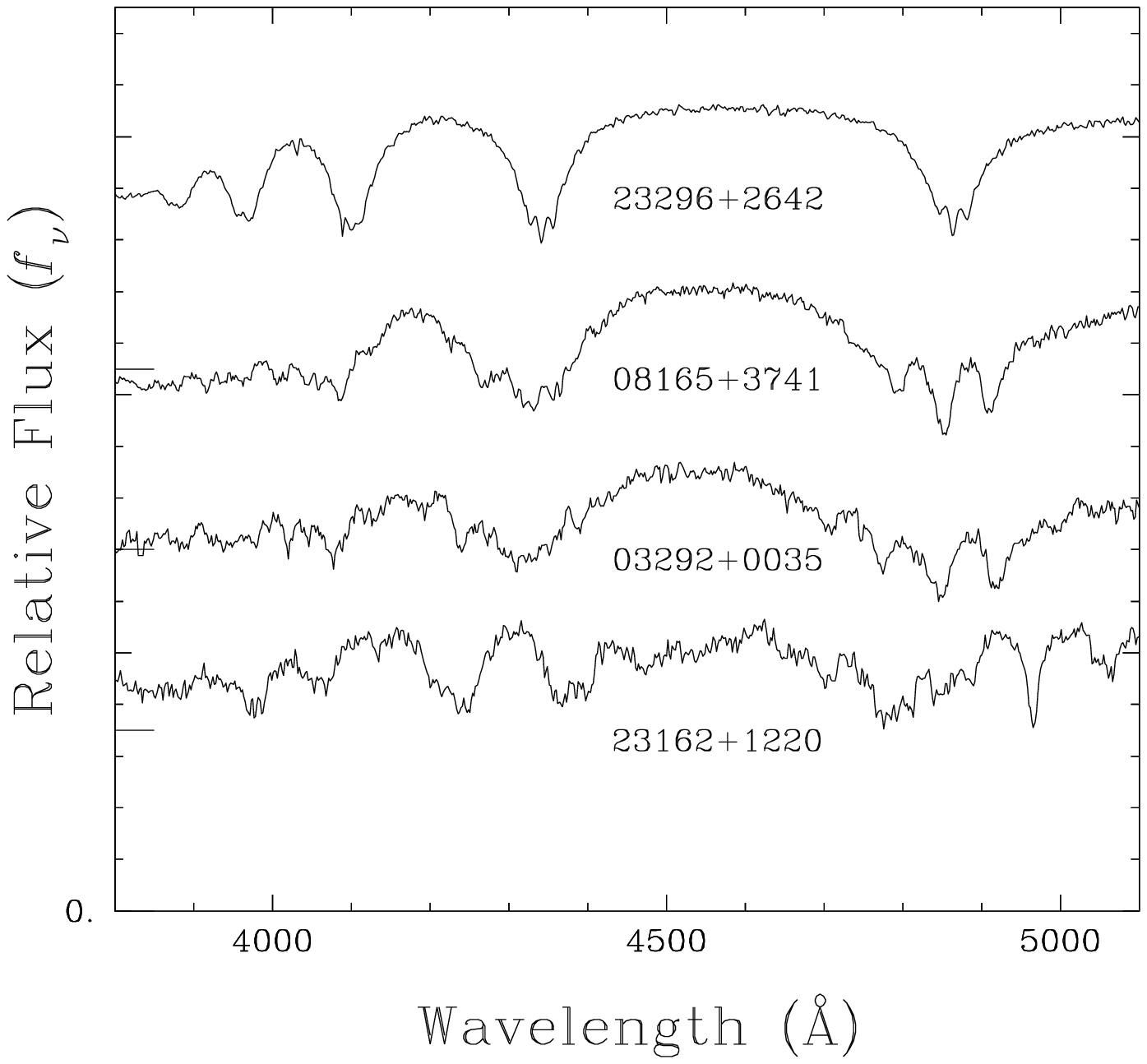]
{Same as Figure \ref{fg:f2} but for the magnetic white
dwarfs in the Kiso survey; these all have hydrogen-rich
atmospheres. The strength of the magnetic field increases from top
($B\sim2.3$~MG) to bottom ($B\sim29$~MG, see references in Table 3).
\label{fg:f4}}

\figcaption[f5]
{Our best fits to the optical spectra of the three DAB stars in our
sample with composite DA+DB models. The atmospheric parameters for
each solution are given in the figure. Both the observed and
theoretical spectra are normalized to a continuum set to unity and the
spectra are shifted from each other for clarity. The detailed analysis
of KUV 02196+2816 has already been published in \citet{lim09} while
the details for KUV 03399+0015 and KUV 14197+2514 will be presented
elsewhere.
\label{fg:f5}}

\figcaption[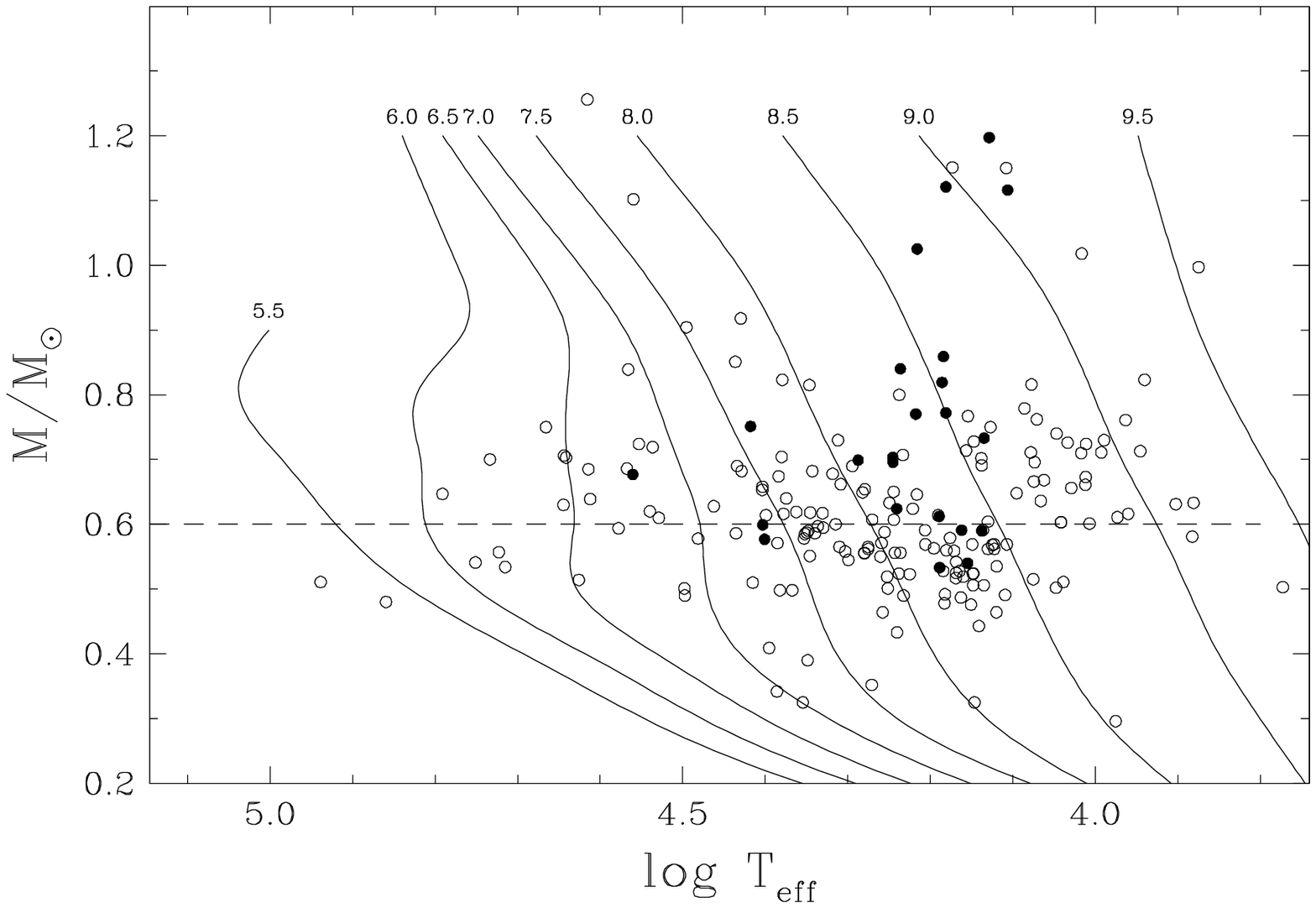]
{Masses of all DA ({\it open circles}) and DB ({\it filled circles})
stars in the Kiso survey as a function of effective temperature, together with
theoretical isochrones labeled in units of log $\tau$, where $\tau$ is
the white dwarf cooling age in years.
\label{fg:f6}}

\figcaption[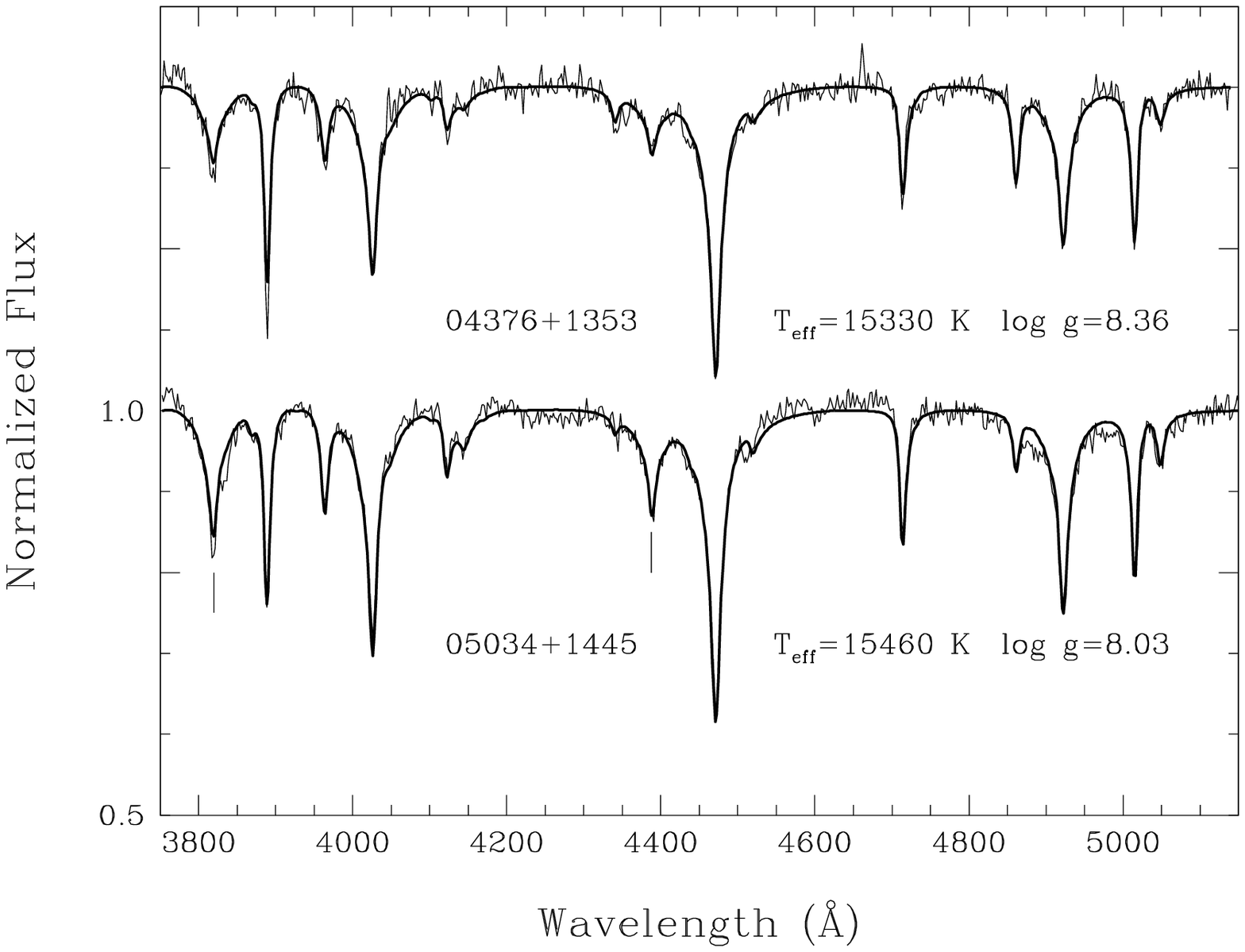]
{Our best fits to two DB stars near 15,000~K, but with significantly
different values of $\logg$ (given in the figure). The spectra are
normalized to a continuum set to unity and the best fitting models are
shown as a thick line. The tick marks indicate the location of the
He~\textsc{i} $\lambda3820$ and $\lambda4388$ lines, which
are the most gravity sensitive in this range of temperature.
\label{fg:f7}}

\figcaption[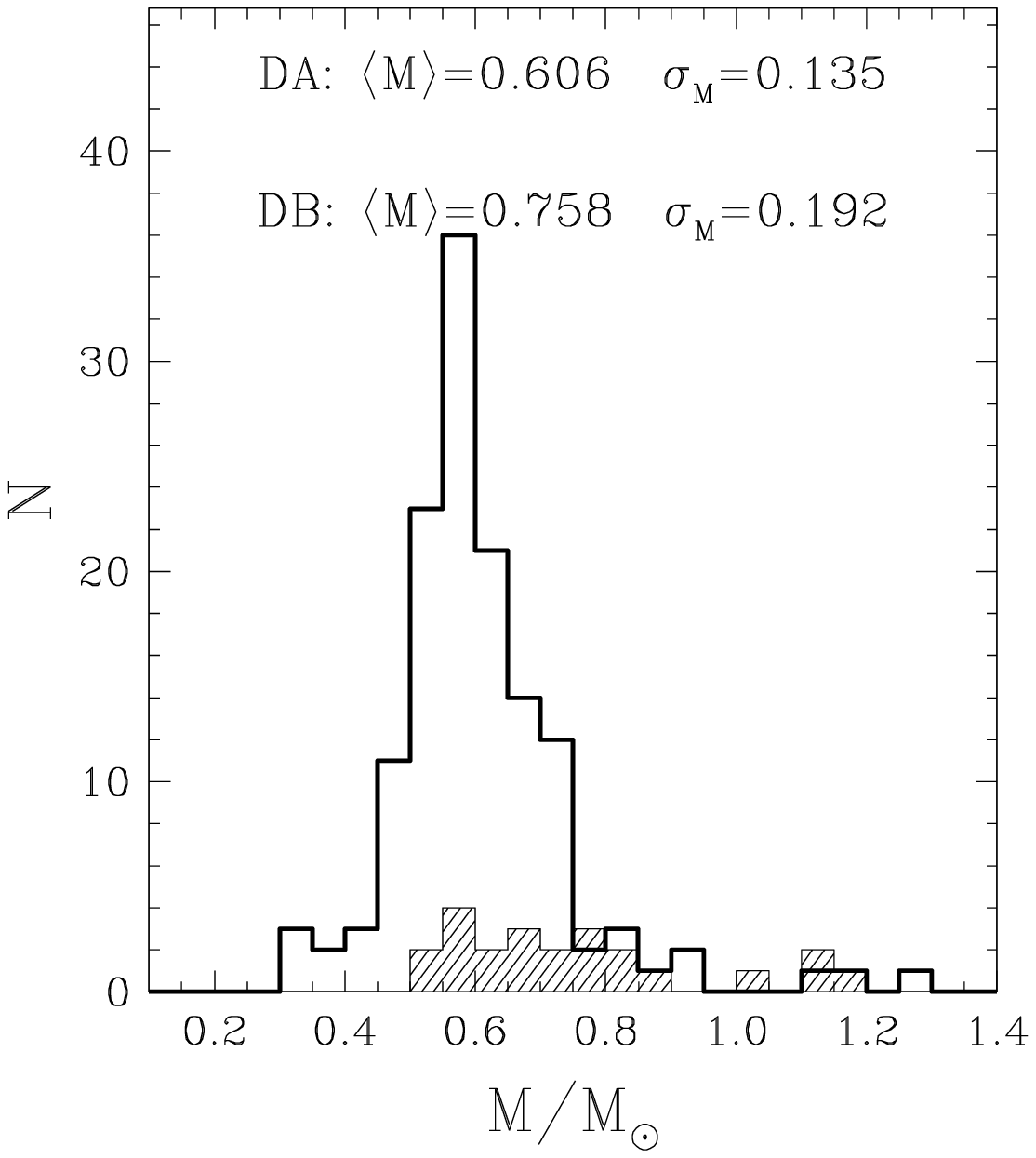]
{Mass distributions for the 136 DA stars in the KUV sample with
$\Te>13,000$~K and the 23 DB white dwarfs.  The masses of DA stars
below this value may be biased, as explained in the text. Mean values
and standard deviations of both distributions are given in the figure.
\label{fg:f8}}

\figcaption[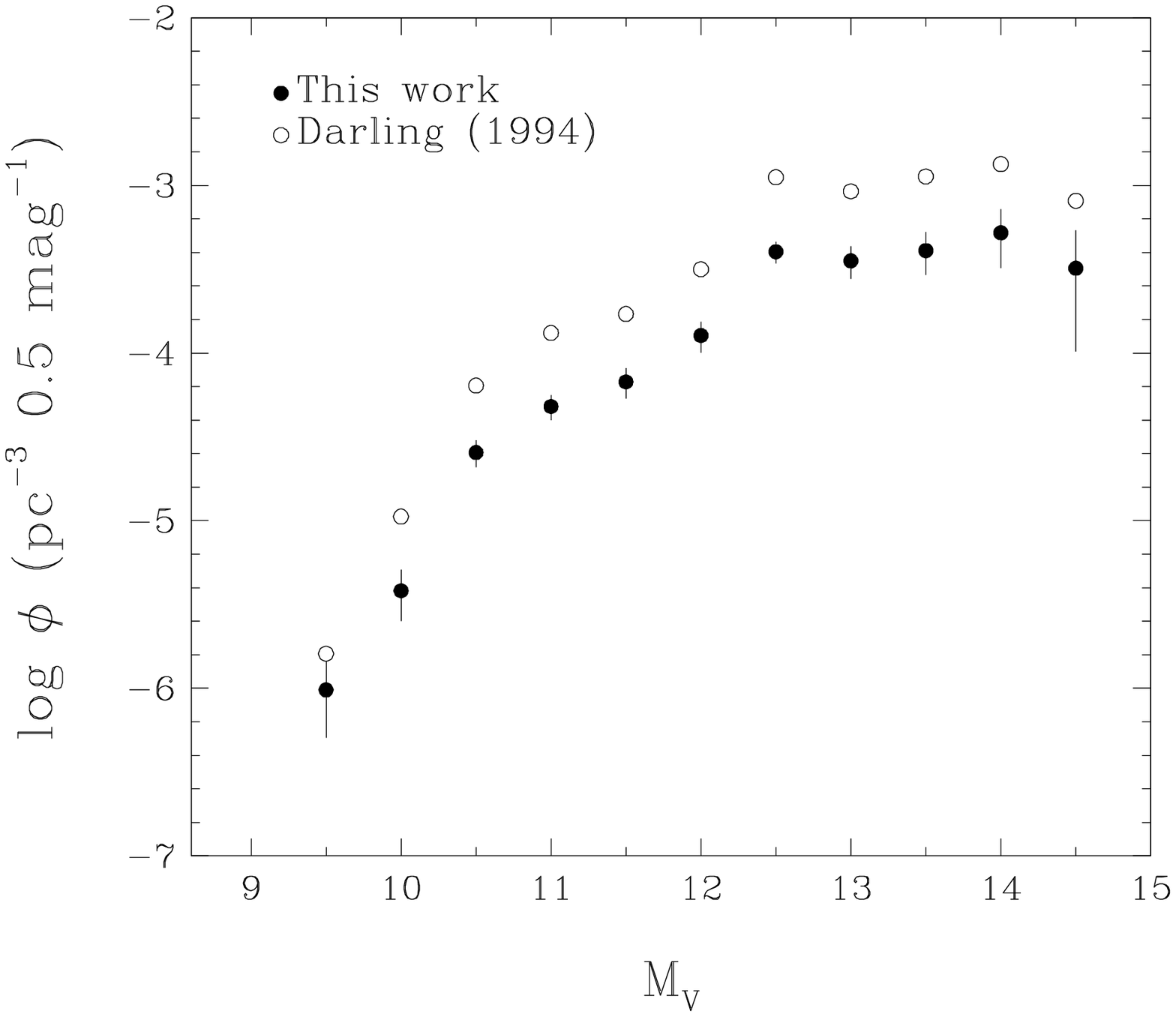]
{Comparison of luminosity functions of white dwarf stars from the Kiso
survey, all spectral types included. The open circles represent the
results published in Table 4.5 of \citet{darling94}, while the filled
circles correspond to our attempt at reproducing his results using the
same input data and method of calculation.
\label{fg:f9}}

\figcaption[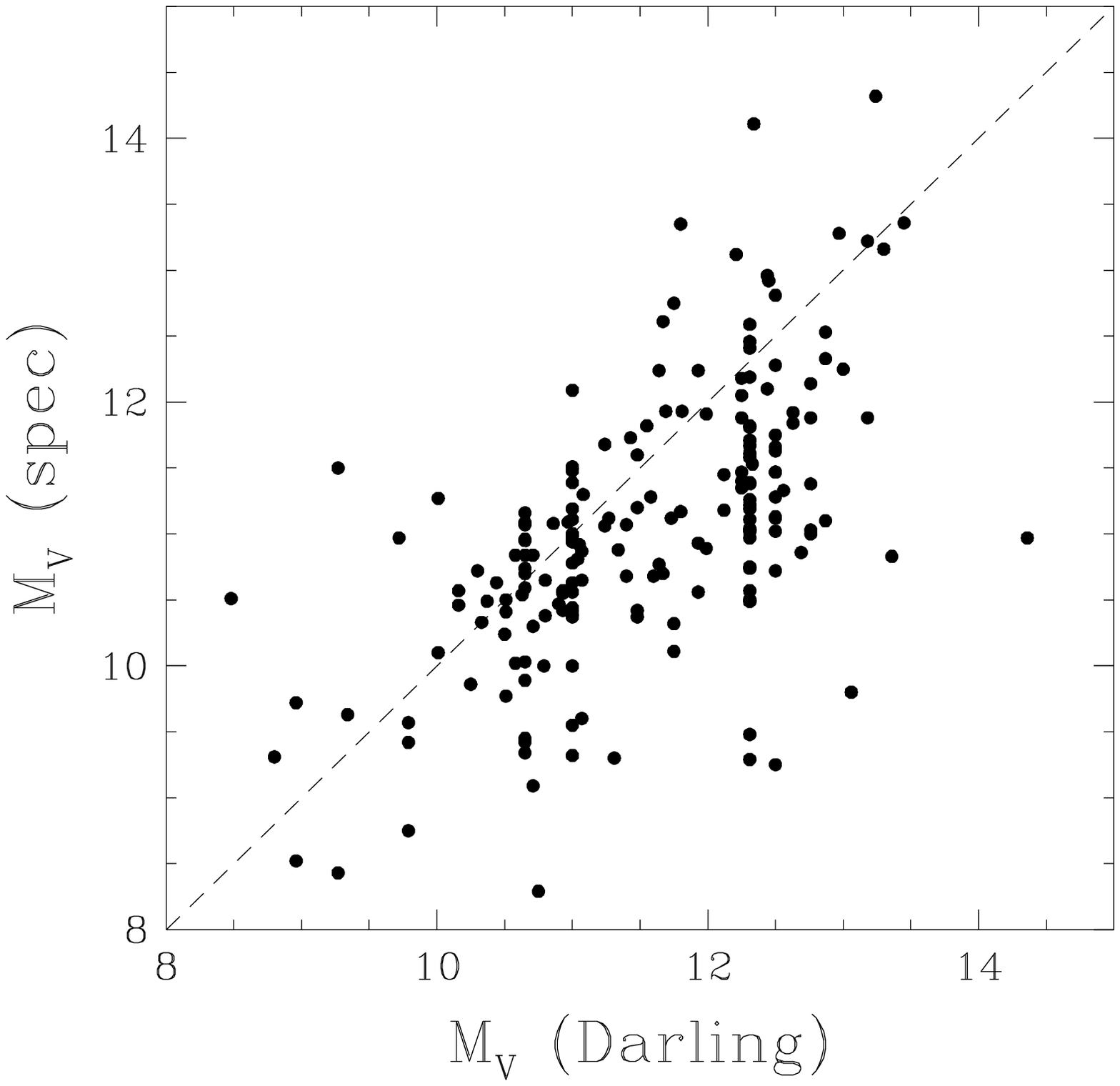]
{Comparison of the absolute visual magnitudes 
obtained from the empirical photometric calibration of
\citet{darling94} and from the spectroscopic method, for all DA 
and DB stars in common.
\label{fg:f10}}

\figcaption[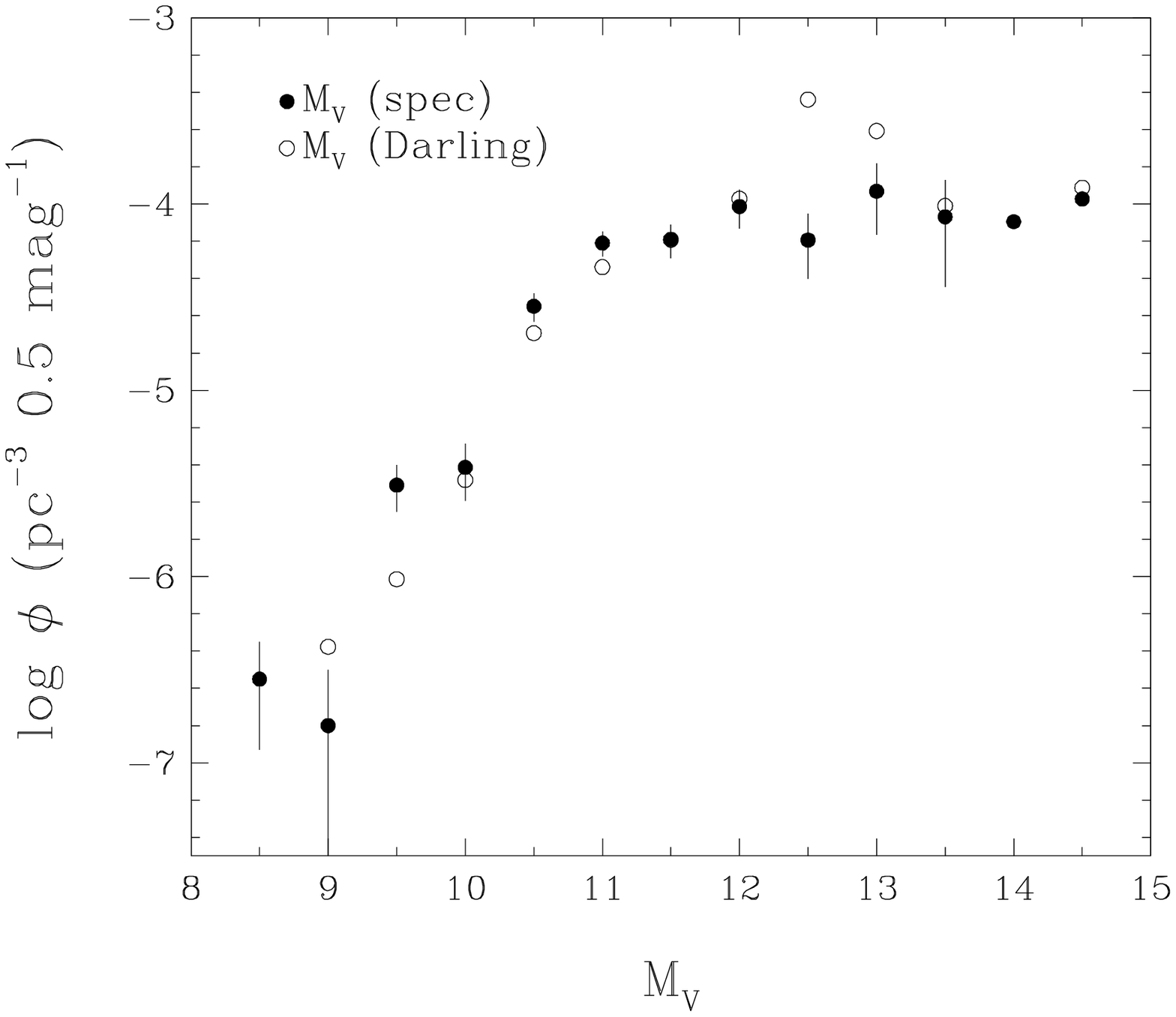]
{Luminosity functions for the DA and DB stars in the Kiso survey that
are in common between our sample and that of \citet{darling94}. The
open circles correspond to the luminosity function calculated using
the approximate $\mv$ values provided in Table 4.3 of
\citet{darling94}, while the filled circles make use of the
spectroscopic $\mv$ values given in our Tables 3 and 4. The magnitude
bin at 14.0 in Darling's data is empty since all cool DQ, DZ, and DC
stars are excluded in this comparison. Also, the last two bins in our
spectroscopic determination contain only one star each, and the
corresponding error bars are too large and not shown here.
\label{fg:f11}}

\figcaption[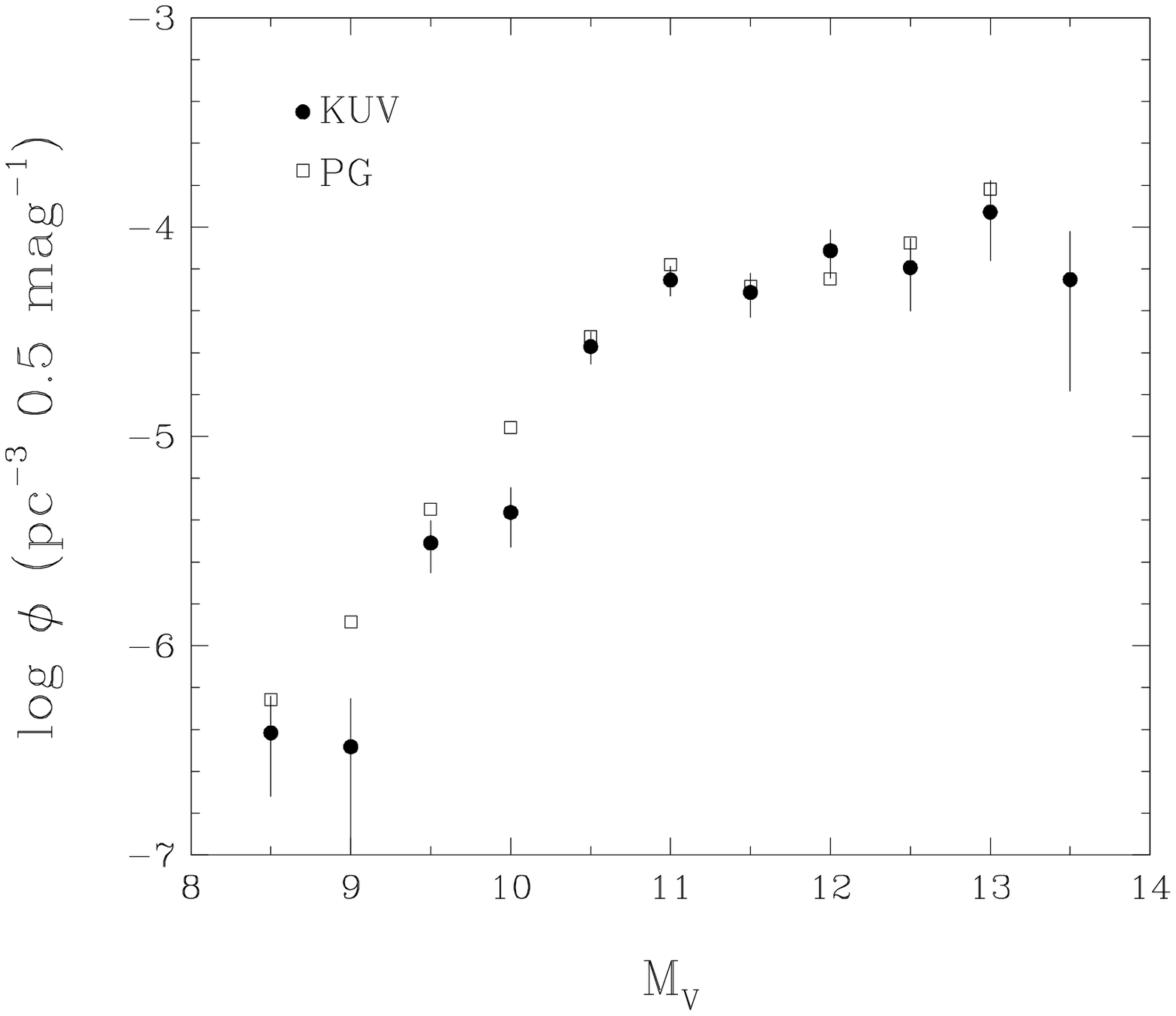]
{Comparison of the luminosity functions for DA white dwarfs in the
Kiso survey (149 stars; this work) and in the PG survey (348 stars;
LBH05).
\label{fg:f12}}

\figcaption[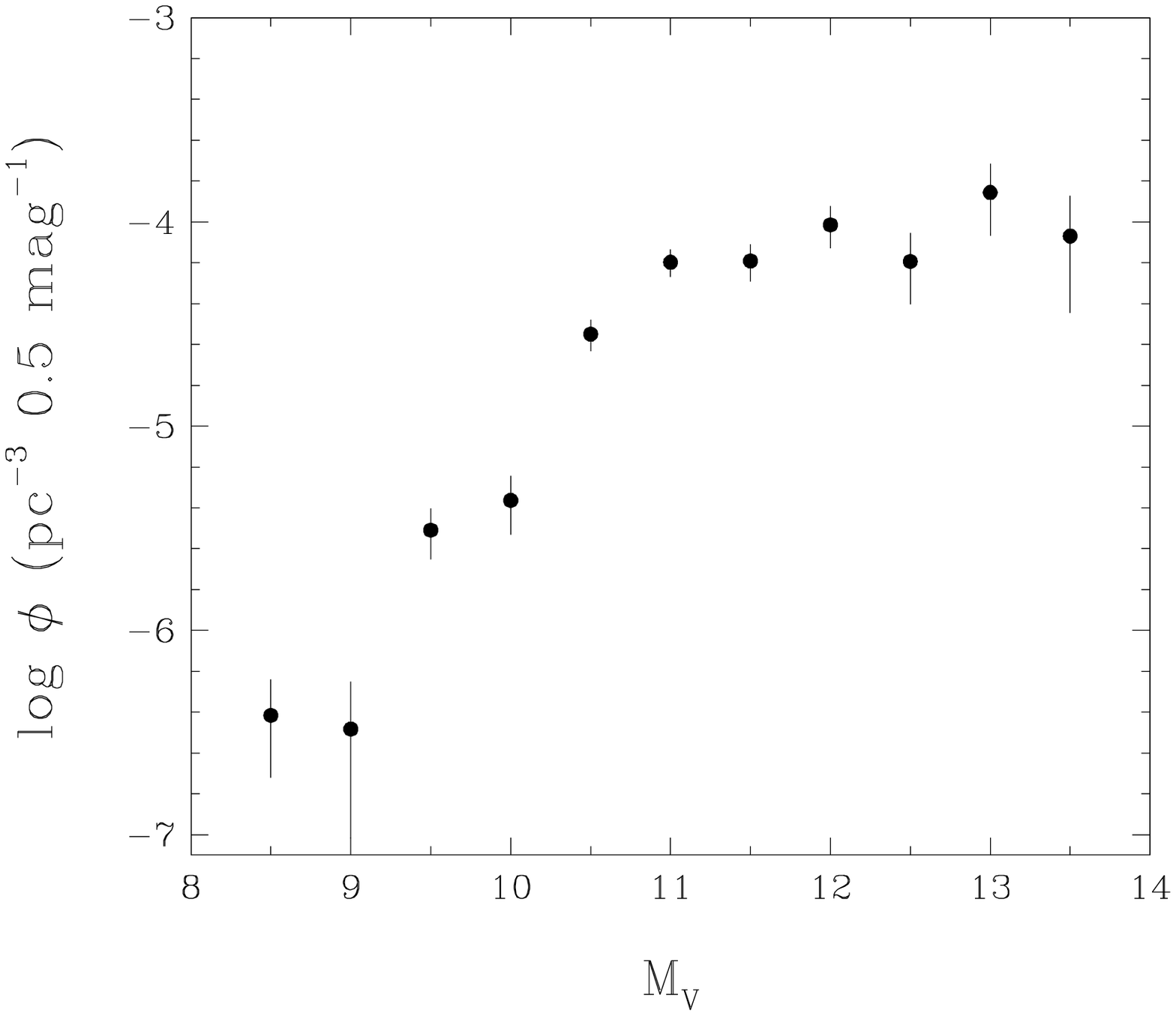]
{The luminosity function for the complete sample of 168 white dwarfs
(149 DA and 19 DB stars) found in the Kiso survey and based on our
spectroscopic $\mv$ values.
\label{fg:f13}}

\figcaption[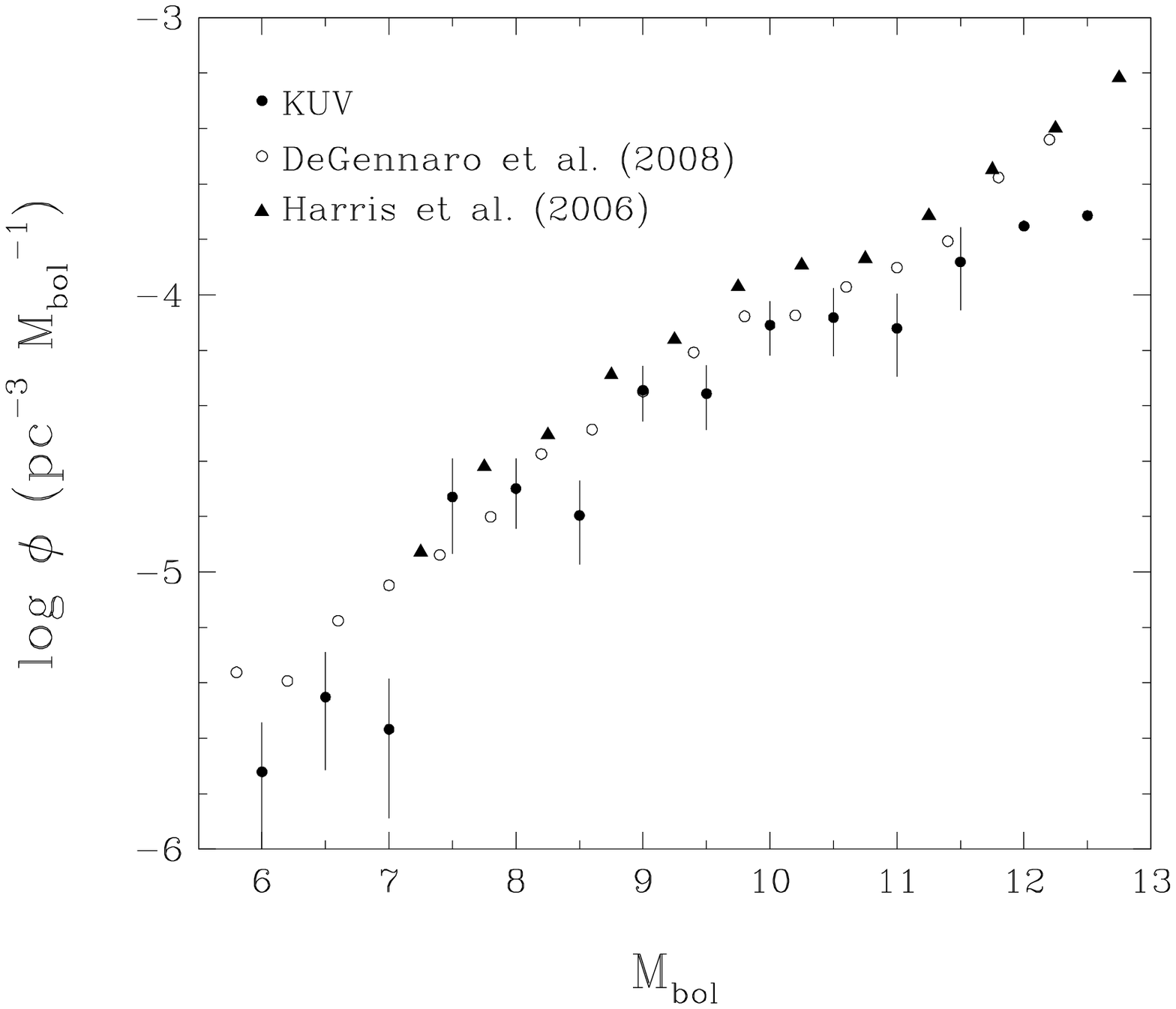] {Our luminosity function for the complete sample
of white dwarfs found in the Kiso survey plotted as a function of
$M_{\rm bol}$ and compared to the luminosity functions obtained by
\citet{harris06} and \citet{degen08} for white dwarfs in the SDSS;
the corresponding uncertainties are roughly equal to the size of
the symbols used here and are thus not shown.
\label{fg:f14}}

\clearpage
\begin{figure}[p]
\plotone{f1.eps}
\begin{flushright}
Figure \ref{fg:f1}
\end{flushright}
\end{figure}

\clearpage
\begin{figure}
\plotone{f2.eps}
\begin{flushright}
Figure \ref{fg:f2}
\end{flushright}
\end{figure}

\clearpage
\begin{figure}
\plotone{f3.eps}
\begin{flushright}
Figure \ref{fg:f3}
\end{flushright}
\end{figure}

\clearpage
\begin{figure}
\plotone{f4.eps}
\begin{flushright}
Figure \ref{fg:f4}
\end{flushright}
\end{figure}

\clearpage
\begin{figure}
\plotone{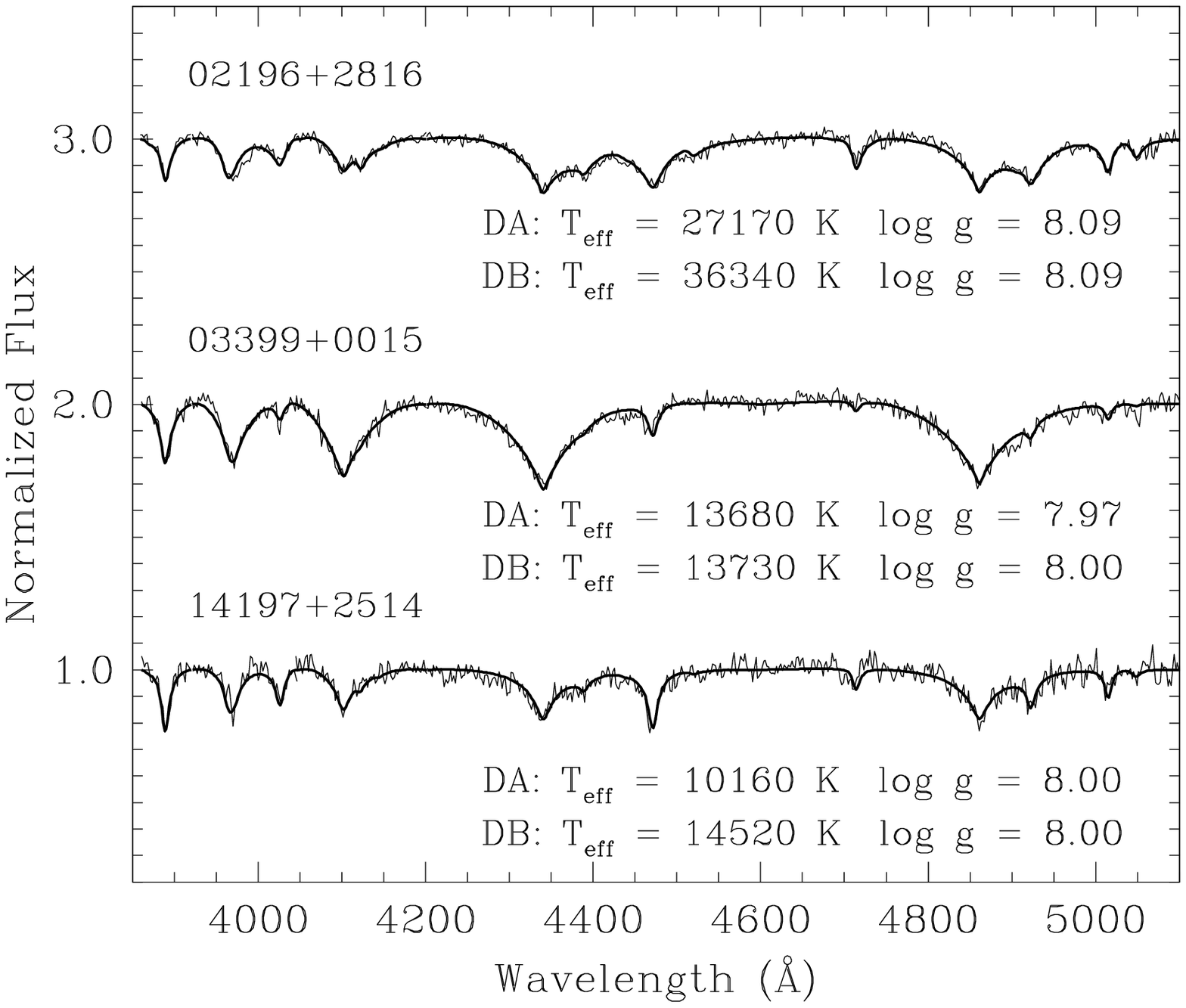}
\begin{flushright}
Figure \ref{fg:f5}
\end{flushright}
\end{figure}

\clearpage
\begin{figure}
\plotone{f6.eps}
\begin{flushright}
Figure \ref{fg:f6}
\end{flushright}
\end{figure}

\clearpage
\begin{figure}
\plotone{f7.eps}
\begin{flushright}
Figure \ref{fg:f7}
\end{flushright}
\end{figure}

\clearpage
\begin{figure}
\plotone{f8.eps}
\begin{flushright}
Figure \ref{fg:f8}
\end{flushright}
\end{figure}

\clearpage
\begin{figure}
\plotone{f9.eps}
\begin{flushright}
Figure \ref{fg:f9}
\end{flushright}
\end{figure}

\clearpage
\begin{figure}
\plotone{f10.eps}
\begin{flushright}
Figure \ref{fg:f10}
\end{flushright}
\end{figure}

\clearpage
\begin{figure}
\plotone{f11.eps}
\begin{flushright}
Figure \ref{fg:f11}
\end{flushright}
\end{figure}

\clearpage
\begin{figure}
\plotone{f12.eps}
\begin{flushright}
Figure \ref{fg:f12}
\end{flushright}
\end{figure}

\clearpage
\begin{figure}
\plotone{f13.eps}
\begin{flushright}
Figure \ref{fg:f13}
\end{flushright}
\end{figure}

\clearpage
\begin{figure}
\plotone{f14.eps}
\begin{flushright}
Figure \ref{fg:f14}
\end{flushright}
\end{figure}

\end{document}